\documentclass[a4paper,11pt]{article}
\pdfoutput=1 

\usepackage{jcappub} 

\usepackage[T1]{fontenc} 
\usepackage{orcidlink}

\usepackage{siunitx}
\DeclareSIUnit{\belm}{Bm}
\DeclareSIUnit{\beli}{Bi}
\DeclareSIUnit{\dBm}{dBm}

\usepackage[super]{nth}
\usepackage{xcolor}
\usepackage{float}

\usepackage{lineno}
\usepackage{amsmath, bm}

\newcommand{\Bgauss}{B_{\mathrm{gauss}}}
\newcommand{\Bouter}{B_{\mathrm{outer}}}
\newcommand{\Bouterm}{B_{\mathrm{outer,\,matched}}}

\newcommand{\Bcore}{B_{\mathrm{core}}}

\newcommand{\Bfinal}{B_{\mathrm{final}}}
\newcommand{\thj}{\theta_{j}}
\newcommand{\ths}{\theta_{s}}
\newcommand{\tha}{\theta_{a}}
\newcommand{\thb}{\theta_{b}}

\newcommand{\RSSlog}{\mathrm{RSS}_{\log}}

\title{The Simons Observatory: Frequency-resolved beam calibration for the SO SATs with POLOCALC}

\author[1]{Nadia Dachlythra\orcidlink{0009-0006-7382-1434},}
\author[1,2]{Gabriele Coppi\orcidlink{0000-0002-6362-6524},}
\author[1]{Ludovico Bizzarri\orcidlink{0009-0008-7401-9229},}
\author[1,2]{Federico Nati\orcidlink{0000-0002-8307-5088},}
\author[1]{Alessandro Novelli\orcidlink{0009-0005-8619-2377},}
\author[3,4]{Elle C. Shaw\orcidlink{0000-0001-5644-8750},}
\author[5]{Samuel Day-Weiss\orcidlink{0009-0003-5814-2087},}

\author[1]{Federico Astori\orcidlink{0009-0009-0890-5895},}
\author[5]{Susanna Azzoni\orcidlink{0000-0002-8132-4896}}
\author[6]{Dillon J. Bass\orcidlink{0009-0001-1448-315X},}
\author[7]{Artem Basyrov\orcidlink{0000-0002-4365-4405},}
\author[1]{Federico Cacciotti\orcidlink{0009-0007-6324-479X},}
\author[8]{Rolando Dünner Planella\orcidlink{0000-0003-3892-1860},}
\author[3,4]{Nicholas Galitzki\orcidlink{0000-0001-7225-6679},}
\author[9]{Remington Gerras\orcidlink{0009-0009-0876-9168},}
\author[10]{John C. Groh\orcidlink{0000-0001-9880-3634},}
\author[11,12]{Jon E. Gudmundsson\orcidlink{0000-0003-1760-0355},}
\author[5]{Saianeesh K Haridas\orcidlink{0000-0001-6519-502X},}
\author[13,14]{Kathleen Harrington\orcidlink{0000-0003-1248-9563},}
\author[8]{Carlos Hervías-Caimapo\orcidlink{0000-0002-4765-3426},}
\author[6]{Bradley R. Johnson\orcidlink{0000-0002-6898-8938},}
\author[15]{Brian Keating\orcidlink{0000-0003-3118-5514},}
\author[16,10]{Adrian T. Lee\orcidlink{0000-0003-3106-3218},}
\author[15]{Anto I. Lonappan\orcidlink{0000-0003-1200-9179},}
\author[1]{Noemi Mezzanzanica\orcidlink{0009-0005-4721-0712},}
\author[5]{Lyman Page\orcidlink{0000-0002-9828-3525},}
\author[1]{Giulia Rancati Cattaneo\orcidlink{0009-0005-4687-1500},}
\author[17,18]{Sara M. Simon\orcidlink{0000-0001-9221-7802}}
\author[19]{Thierry Souverin\orcidlink{0009-0004-1256-0491},}
\author[5,20]{Yoshinori Sueno\orcidlink{0000-0002-3644-2009},}
\author[21]{Edward J. Wollack\orcidlink{0000-0002-7567-4451},}
\author[5]{Kyohei Yamada\orcidlink{0000-0003-0221-2130},}
\author[1,2]{Mario Zannoni\orcidlink{0000-0002-4495-571X}.}

\affiliation[1]{Department of Physics, University of Milano-Bicocca, Piazza della Scienza 3, 20126, Milano, Italy}
\affiliation[2]{Istituto Nazionale di Fisica Nucleare, INFN, Sezione Milano-Bicocca, Piazza della Scienza 3, 20126, Milano, Italy}
\affiliation[3]{Department of Physics, University of Texas at Austin, Austin, TX 78712, USA}
\affiliation[4]{Weinberg Institute for Theoretical Physics and Texas Center for Cosmology and Astroparticle Physics, The University of Texas at Austin, Austin, TX 78712, USA}
\affiliation[5]{Joseph Henry Laboratories of Physics, Jadwin Hall, Princeton University, Princeton, NJ, USA 08544}
\affiliation[6]{Department of Astronomy, The University of Virginia, Charlottesville, VA 22904, USA}
\affiliation[7]{Astroparticule et Cosmologie (APC), Université Paris Cité, CNRS, 75013 Paris, France}
\affiliation[8]{Instituto de Astrofísica and Centro de Astro-Ingeniería, Facultad de Física, Pontificia Universidad Católica de Chile, Av. Vicuña Mackenna 4860, 7820436, Macul, Santiago, Chile}
\affiliation[9]{Department of Physics, University of Southern California, Los Angeles, CA 90089, USA}
\affiliation[10]{Physics Division, Lawrence Berkeley National Lab, Berkeley, CA 94720 USA}
\affiliation[11]{Science school, University of Iceland, Sæmundargata 2, 102 Reykjavík, Iceland}
\affiliation[12]{The Oskar Klein Centre, Department of Physics, Stockholm University, SE-106 91 Stockholm, Sweden}
\affiliation[13]{High Energy Physics Division, Argonne National Laboratory, Lemont, IL 60439, USA}
\affiliation[14]{Department of Astronomy and Astrophysics, University of Chicago, Chicago, IL, 60637, USA}
\affiliation[15]{Department of Physics, University of California, San Diego, La Jolla, California 92093}
\affiliation[16]{Department of Physics, University of California, Berkeley, CA 94720 USA}
\affiliation[17]{Department of Astronomy and Astrophysics, University of Chicago, Chicago, IL, 60637, USA}
\affiliation[18]{Cosmic Frontier Division, Fermi National Accelerator Laboratory, Batavia, IL, 60510, USA}
\affiliation[19]{LAPP - Laboratoire d'Annecy de Physique des Particules (9 Chemin de Bellevue - BP 110 - 74941 Annecy-le-Vieux CEDEX - France - France}
\affiliation[20]{Kavli Institute for the Physics and Mathematics of the Universe (Kavli IPMU, WPI), UTIAS, The University of Tokyo, Kashiwa, Chiba 277-8583, Japan}
\affiliation[21]{NASA Goddard Space Flight Center, 8800 Greenbelt Road, Greenbelt, MD 20771, USA}

\emailAdd{konstantina.dachlythra@unimib.it}

\abstract{We present a far-field beam calibration of the Simons Observatory (SO) Small
Aperture Telescopes (SATs) using scanned single-frequency measurements of a
drone-mounted source emitting in the \SI{90}{\giga\hertz} band, where the SATs
have an angular resolution of ${\sim}\SI{30}{\arcminute}$. We then compare the
results from two campaigns in April and December 2025 to the band-averaged
beams from planet observations by the SATs. Measuring the telescope beam response at different frequencies across the observing bands enables modeling the system's frequency-dependent behavior. This helps track optical non-idealities. The comparison with planet beams informs the frequency resolution necessary to accurately reconstruct the band-averaged beams in this way; it uses only the average drone and planet products, as it is intended to demonstrate the feasibility of the frequency-resolved calibration scheme rather than to set the precise beam covariance and instrument requirements. We find that the beams of the two SATs near \SI{90}{\giga\hertz} follow the frequency scaling expected for diffraction-limited optics, with only modest deviations observed in the nearest lobes of one telescope. We establish consistency between the real band-averaged planet beams and the synthesized drone band-averaged beams for the center wafers of both telescopes, with an average deviation smaller than $4\%$ for multipoles $\ell<500$.}

\begin{document}
\maketitle
\flushbottom

\section{Introduction}
\label{sec:intro}

The Cosmic Microwave Background (CMB) is a cornerstone of modern cosmology, offering a snapshot of the Universe at recombination and encoding detailed information about its origin and evolution. Beyond temperature anisotropies, the CMB exhibits a faint linear polarization field that can be decomposed into parity-even, curl-free $E$-mode and parity-odd, divergence-free $B$-mode components, each tracing different physical processes. In the standard inflationary paradigm, primordial fluctuations generated during an early phase of accelerated expansion, give rise to both scalar and tensor perturbations in the spacetime metric. Scalar (density) perturbations produce only $E$-mode polarization at linear order, while tensor perturbations—primordial gravitational waves—generate both $E$- and $B$-modes. The presence of $B$-mode polarization therefore provides a distinctive signature of tensor modes and a direct probe of inflationary physics (e.g.~\citep{SeljakZaldarriaga1997, HuWhite1997}). The relative amplitude of tensor-to-scalar perturbations is commonly parametrized by the tensor-to-scalar ratio, $r$, defined through the ratio of their respective power spectra. Constraining $r$ through precise measurements of CMB polarization is a central objective of current experiments, as it directly informs the energy scale and dynamics of inflation (e.g.~\citep{Lyth_1997, Baumann2009, Planck2020CosmoParams}).

Achieving the scientific objectives of a CMB instrument depends critically on
comprehensive understanding of its response and systematic effects
\citep{Hu2003, Shimon2008}. Imperfect optics and miscalibration can induce
temperature-to-polarization ($T$-to-$P$) leakage and mixing of the two
polarization components ($E$-to-$B$), contaminating the faint primordial
$B$-mode signal. To mitigate these effects, experiments devote substantial effort to calibrating the instrument's gain, pointing, polarization angle, and spatial response to a point source \citep{Abitbol2021}. This spatial response, commonly called the point-spread function (PSF) or beam, is usually characterized through in-situ calibration observations; its squared harmonic transform (the beam window function), expressed as a function of multipole, is then deconvolved from the sky spectra prior to component-separation analysis. Note that beam convolution acts as a multiplicative filter in harmonic space: a purely scale-dependent suppression of power that mixes neither temperature with polarization nor the $E$- and $B$-modes. Any leakage will only appear from non-idealities not captured by the window function.

Although artificial sources have been successfully employed in the field
\citep{polarbear_2014_subdegree, bicep2_2014_experiment} and are planned for future generations
\citep{Ritacco_2024, 2026_leocalsat}, calibration most commonly relies on bright
celestial sources such as Jupiter or Saturn, and in some cases fainter planets or
the Moon, depending on detector dynamic range and saturation limits
\citep{calibration_wmap_2011, Hasselfield_2013, planck_planetflux_2017,
Datta_2024}. These astrophysical calibrators are often well approximated as grey
bodies over CMB observing bands, which makes them especially useful for
band-averaged calibration. However, all the instrument properties of interest are known to vary
non-negligibly, and not always trivially, within the wide observing bands of CMB
telescopes. If the telescope is equipped with more complex optical components,
such as polarization modulators, these introduce additional frequency
dependence \citep{Takakura_2017, Duivenvoorden2021}. Left unmodeled, such
variations can interfere with the frequency-dependent assumptions used to
separate the CMB from Galactic foregrounds based on their spectral energy
distributions (SEDs), biasing the recovered foreground and cosmological
parameters. This raises the question of whether a frequency-resolved calibration scheme may provide advantages over a purely
band-averaged treatment.

A representative example of the telescope's chromatic behavior is its beam, which varies naturally across the telescope's observing bands. The interplay between beam chromaticity and the sky becomes significant once the beam is convolved with sky components whose emission carries distinct spectral signatures \citep{Giardello_2024, Dachlythra_2025}. Because each component is weighted by the beam's frequency response in a different way, the effective beam imprinted on the data couples to the spectral behavior of every component, and this coupling propagates into the recovered foreground parameters and may ultimately bias the constraint on the tensor-to-scalar ratio, $r$ \citep{Dachlythra_2025}. Narrow-band sources such as those described in \citep{Coppi:2025} are ideal calibrators for the frequency-resolved beam, provided they can be placed in the far field of the telescope for a sufficiently long time. Measuring the beam at a discrete set of frequencies enables the level of chromaticity to be quantified and identify departures from the expected optical behavior. Where needed, this enables scaling corrections to be applied in multipole space before the sky spectra are used in the cosmological analysis. Moreover, with a careful choice of frequencies and good knowledge of both the instrument bandpass and the source power, these single-frequency measurements can be combined to closely approximate the band-averaged beam.

In this paper, we present a frequency-resolved beam calibration scheme based on measurements taken with a thermal narrow-band source mounted on a drone \citep{Nati_2017, Dunner_2021, Coppi:2025} and flown through the beam of the Simons Observatory (SO) Small Aperture Telescopes (SATs). The SATs will survey the polarized microwave sky from the Atacama Desert in Chile across six frequency bands spanning roughly \SIrange{27}{280}{\giga\hertz}, with the goal of measuring $r$ to a statistical precision of $\sigma(r) = 0.003$ with the first three SATs over five years of observation \citep{SO_overview_2019}, improving to $\sigma(r) \leq 0.0012$ once three more are deployed and ten years of data are accumulated \citep{Abril-Cabezas_2026}. By contrast, the SO Large Aperture Telescope (LAT) cannot be calibrated in this way, as this would require flying the source $\sim$ \SI{20}{\kilo\meter} away from the telescope. The drone-borne calibration effort was initiated through two projects, POLOCALC \citep{Nati_2017} and HoverCal \citep{Dunner_2021}, aiming to calibrate the polarization angle of CMB telescopes with \SI{0.1}{\degree} accuracy in the 90 and \SI{150}{\giga\hertz} bands. The drone-borne sources carry fully polarized, well-controlled signals, making them uniquely suited to characterize both the polarized beam and the polarization angle of the SATs, a measurement crucial to the success of their scientific mission. Though $B$-modes peak near $\ell = 80$, we model the SAT beam out to $\ell = 500$ (see Section~\ref{sec:planet_comparison}) to improve component separation and enable future joint constraints with the SO LAT. While planet observations provide the most robust band-averaged beam measurements, to be presented in a forthcoming paper \citep{Day-Weiss:2026}, the drone-based approach uniquely resolves the beam as a function of frequency. Because the SAT bands are wide, reaching fractional bandwidths of order $\sim25\%$, the frequency dependence of the beam across each band must be carefully modeled if its imprint on the large-scale $B$-mode spectrum is to be controlled. This paper focuses on the chromaticity of the \SI{90}{\giga\hertz} band extracted from POLOCALC results, while those from HoverCal will be presented in future work. Sources emitting in the \SI{220}{} and \SI{280}{\giga\hertz} bands have also been acquired and are planned for deployment in future campaigns.

The paper is organized as follows. Section \ref{sec:system_overview} summarizes the drone hardware and the SAT instrument configuration. Section \ref{sec:measurements} covers the preparations and tests for a calibration campaign, and Section \ref{sec:analysis} describes the analysis pipeline applied to the observations. Section \ref{sec:results} presents the beam-fitting pipeline and its results: the variation of the beam across the band and the comparison of the synthesized drone beam with the reconstructed planet band-averaged
beam. Finally, Section \ref{sec:conclusion} summarizes the work and outlines directions for future investigation.

\section{Hardware overview}
\label{sec:system_overview}

\begin{figure}
    \centering   
    \includegraphics[width=0.25\linewidth]{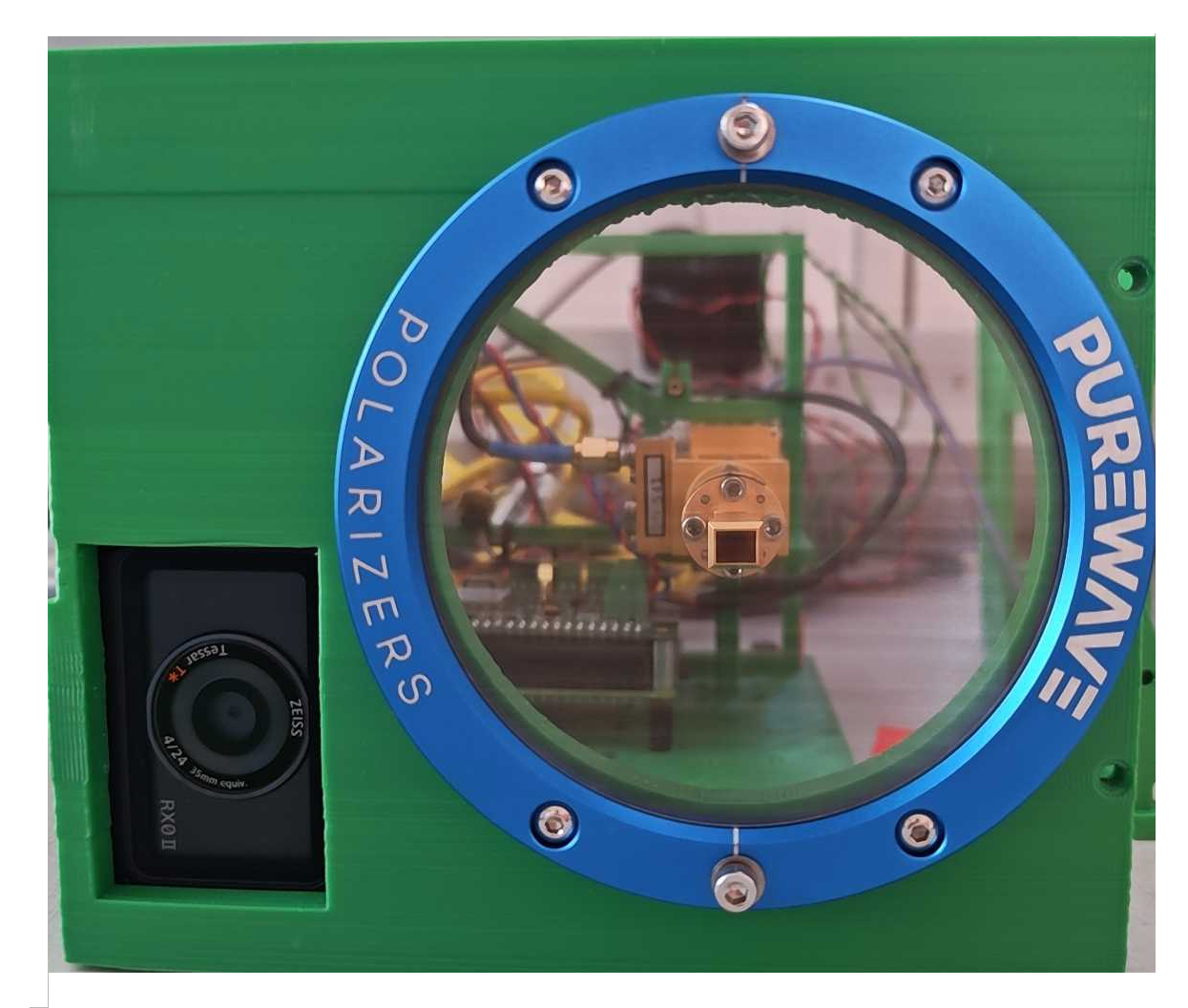}\includegraphics[width=0.76\linewidth]{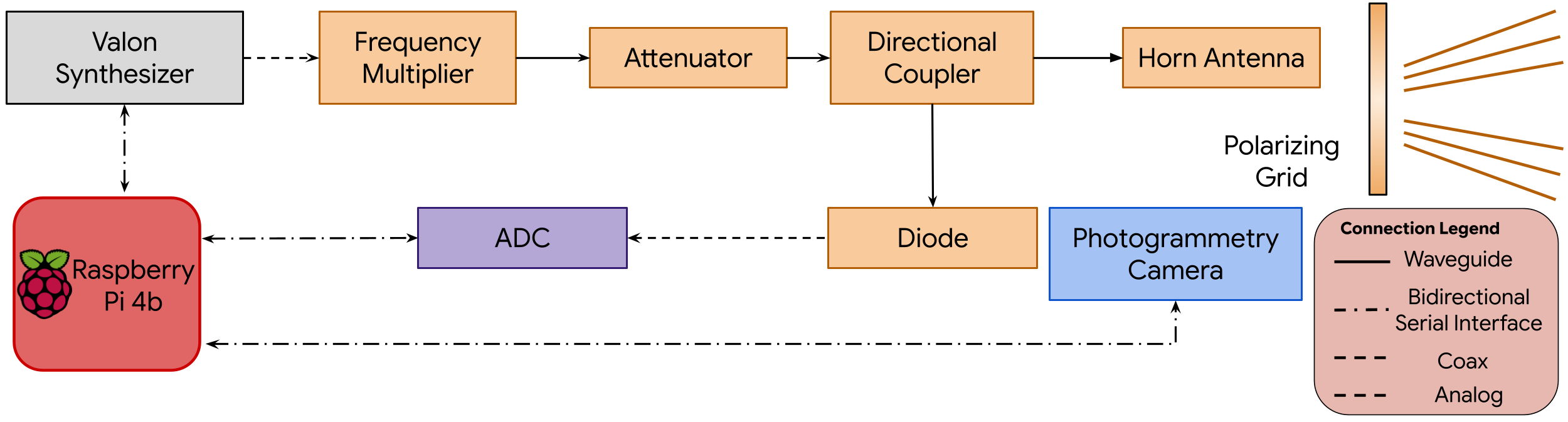}    \caption{\label{fig:drone_setup} Picture (left) and schematic (right) of the main components of the drone setup. The Valon is only included in the schematic.}
\end{figure}

\paragraph{} The calibrator hardware is composed of two main components: the
carrier which is the drone, and the calibration source payload. The
drone is a DJI Matrice 600 Pro capable of carrying payloads up to \SI{6}{\kilo
\gram} and reaching altitudes of up to \SI{500}{\meter} above the SO site before encountering pressure and stabilization limits. The interface between the drone and the scientific payload is a DJI Ronin
MX gimbal that stabilizes the source during flight. The source itself is described
in \cite{Coppi:2025} and consists of a frequency synthesizer that generates a very
narrow-band frequency coupled to a 6x frequency multiplier that boosts the frequency to
the scientific band (from \SI{15}{\giga\hertz} to \SI{90}{\giga\hertz}). The use of
a synthesizer allows us to select the frequency with a high degree of precision
(few tens of \SI{}{\kilo\hertz}). The frequency synthesizer output is square-wave modulated at $f_{\mathrm{chop}}\sim$ \SI{37}{\hertz} by switching the power between \SI{0}{\dBm} and \SI{-14}{\dBm}, the latter low enough to not be multiplied by the frequency multiplier, producing an effective amplitude modulation. The chopping frequency was chosen such that $f_{\mathrm{chop}}<<1/2\pi\tau$, where $\tau$ is the detector time constant. Compared to the system described in \cite{Coppi:2025},
we upgraded to a Virginia Diodes W-Band 6x multiplier that has an internal
attenuator that allows for fine control of the output power. Additionally, in the
December 2025 campaign, we integrated a simple thermostat circuit that keeps the
multiplier at a constant temperature ($\pm$ \SI{1}{\kelvin}), stabilizing the
output power. More details of this new system will be presented in
\cite{Bizzarri_2026prep}. After the multiplier there is a fixed attenuator, a
directional coupler and a relatively low-gain/large-beam antenna \cite{Coppi:2025}.
To define the polarization of the emitted signal, a \SI{10}{\centi\meter}-diameter wire grid is placed at the antenna output; the transmitted signal is polarized perpendicular to its tungsten wires, so the wire orientation sets the polarization direction. Connected to the directional coupler is a
detector diode that monitors the output power and whose signal is recorded to help determine the phase of the chopping signal. The position of the
source is reconstructed to a few centimeter accuracy using Real-Time Kinematics
\citep{Manandhar1999}. Finally, the attitude is reconstructed using a combination
of photogrammetry — based on imagery from an onboard camera — and high speed
inertial sensors such as accelerometers and gyroscopes. Figure \ref{fig:drone_setup} shows a picture and schematic of the calibrator setup. 

The SO SATs are compact, self-contained cryogenic telescopes designed for CMB polarization measurements. Each SAT has a \SI{42}{\centi\meter} optical aperture and a \SI{35}{\degree} field of view, using a refractive optical design with three metamaterial anti-reflection coated silicon lenses housed in an aluminum optics tube cooled to \SI{1}{\kelvin}. The cryogenic system includes nested temperature stages, with an outer \SI{40}{\kelvin} stage, an inner \SI{4}{\kelvin} stage, and a dilution refrigerator that cools the focal plane below \SI{0.1}{\kelvin}. The focal plane contains over 12,000 Transition-Edge-Sensor (TES) bolometers arranged in about 3,000 pixels populating seven hexagonal wafers that form an even larger hexagon \citep{McCarrick_2021, Dutcher:2024}. The center wafer, which is the one employed in this paper, is conventionally named `ws0' while the edge wafers carry names from `ws1' to `ws6'. The exact layout of SAT1 and SAT3 is shown in Figure 10 of \cite{Harrington:2026}. Each SAT observes simultaneously in two low, mid, or ultra-high frequency bands (LF, MF, or UHF bands) roughly centered on (\SI{27}{}, \SI{39}{}), (\SI{93}{}, \SI{145}{}), and (\SI{225}{}, \SI{280}{}) \SI{}{\giga\hertz}, respectively.  Moreover, the polarization is modulated by a continuously rotating half-wave plate (HWP) near the aperture stop, which helps separate the sky signal from atmospheric and instrumental effects \citep{Yamada2024, Sugiyama2024}. SAT1 and SAT3 are the telescopes observing in the MF bands and consequently the ones we target with the drone measurements. While the two telescopes share largely the same design, they also carry a few differences that could shape their beams differently. SAT1 is equipped with a reflective comoving shield and an absorptive forebaffle, whereas SAT3 has only an integrated forebaffle, which in this case is reflective \citep{Harrington:2026}. Moreover, the anti-reflection (AR) coatings differ between the two telescopes: SAT1 uses mullite-duroid on the HWP and all alumina IR filters \citep{Sakaguri:24}, whereas SAT3 uses metamaterial coatings on the HWP alumina and the \SI{4}{\kelvin} alumina IR filter \citep{Golec:22}. A description of the design, testing, and integration of the SATs can be found in \cite{Galitzki_2024}.

\section{Measurements}
\label{sec:measurements}

\paragraph{}For drone measurements at the observatory, we lay down forty colorful circular targets, each
\SI{80}{\centi\meter} in diameter, whose positions are later used in the
photogrammetry. The targets are distributed within a \SI{200}{\meter} radius
circle centered approximately \SI{25}{\meter} east of the SAT3 (roughly matching
the center of the site plateau). Each target is placed \SI{10}{} to \SI{15}{\meter} away from adjacent targets while the overall distribution is random. This is because a regular 3D grid of points, when projected onto a 2D image, would introduce ambiguity in the reconstruction of the camera attitude. The absolute positions of the targets are measured with the GPS with a precision of approximately $\SI{10}{\centi\meter}$ (see Figure 2 of \cite{Coppi:2025}), and their corresponding coordinates on the camera sensor (image space) are extracted as the first step of the photogrammetry pipeline. With the targets' 3D world coordinates and their 2D image-space projections in hand, photogrammetry recovers the six degrees of freedom of the camera (its position and orientation) by solving the system of equations that relates each world point to its location on the sensor. Note that the \SI{10}{\centi\meter} uncertainty refers to the GPS horizontal position and together with a \SI{5}{\centi\meter} vertical error allows for reconstructing the drone azimuth and elevation to within \SI{0.01}{\degree} accuracy \citep{Coppi:2025}. Finally, a few test flights without the \SI{90}{\giga\hertz} source typically take place at the beginning to ensure that the drone body alone does not saturate the detectors.  

\begin{figure}
    \centering
    \includegraphics[width=0.95\linewidth]{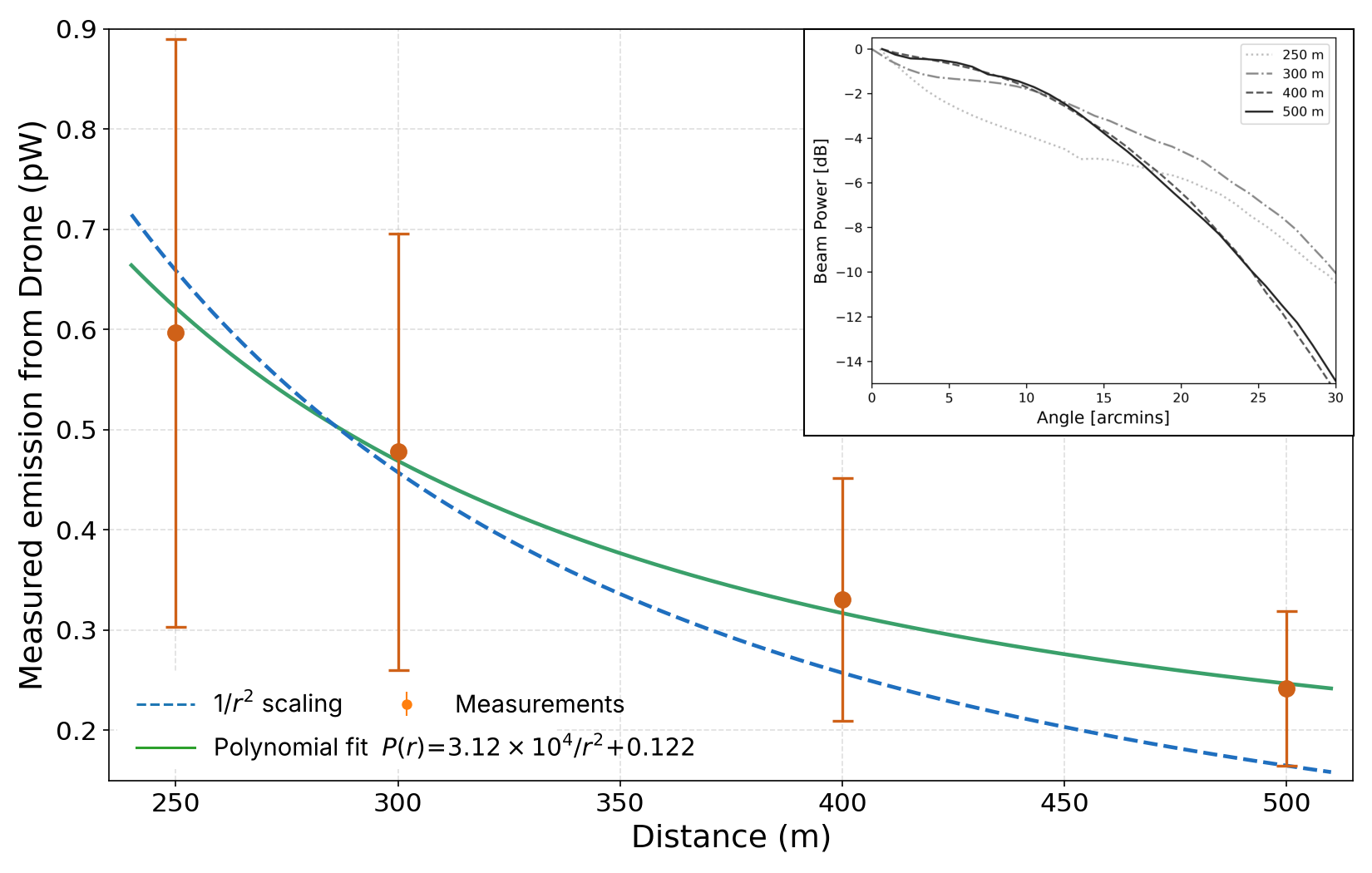}   \caption{\label{fig:Drone_emission_f_disr}Drone emission as measured by SAT1 detectors in the SAT \SI{90}{\giga\hertz} band when flown with its source off as a function of distance from the telescope (red dots and error bars). The flattening of the power at shorter distances, as compared to the expected $1/r^{2}$ scaling (blue dashed curve), and subsequent increase in uncertainty is the result of detector saturation. This is further supported by the inset, which displays the azimuthally averaged beam power as a function of angle for four drone distances: the main lobe is most strongly flattened at \SI{250}{\meter} and \SI{300}{\meter}, while the profiles at \SI{400}{\meter} and \SI{500}{\meter} converge and recover a well-defined shape, indicating that detector saturation subsides at larger distances. The demonstrated measurements are from April 2025. A subsequent campaign in December 2025 confirmed consistency at \SI{400}{\meter} and \SI{500}{\meter}}
\end{figure}

The drone acts as a thermal source of T$\sim$\SI{300}{\kelvin} and needs to be placed at least \SI{400}{\meter} away from the SATs to avoid saturating the \SI{90}{\giga\hertz} detectors \cite{Dutcher:2024}. Figure \ref{fig:Drone_emission_f_disr} shows the drone power as measured by the SAT1 ws0 wafer, when flown without the source at different distances from the telescope (red dots and error bars). All measurements were acquired at distances well beyond the far-field threshold of the SATs ($\sim$\SI{120}{\meter}) at \SI{90}{\giga\hertz} and under consistent atmospheric conditions, with a PWV of approximately \SIrange{1}{1.5}{\milli\meter}. Since atmospheric water vapor attenuates the signal, a PWV correction is normally required to compare measurements taken under different conditions; because our data were acquired under stable, low-PWV conditions, no such correction was applied. The estimated expected power scaling is also shown (blue dashed curve). This is computed by taking the expected power of the drone source as seen by a single detector (estimated in \cite{Coppi:2025}), applying the expected inverse-square ($1/r^{2}$) dependence on distance, and including the average measured detector optical efficiency \citep{Harrington:2026}. The flattening at near distances can be mostly explained as saturation of the detectors. This is further motivated by the substantial increase in the error bars (estimated as the standard deviation of the per-detector power measurements in a single flight) as the drone-telescope distance decreases. As additional proof, the inset in the top right of the same panel presents the main beam of the detectors reconstructed at these four distances, confirming saturation once again through the flattened peaks at \SI{250}{} and \SI{300}{\meter}. Although the beams at both \SI{400}{} and \SI{500}{\meter} form well-defined main lobes, we take the latter as the nominal distance, as it lets us drive the source at higher power and thereby raise the SNR of the chopped signal. Moreover, the source is not a perfect thermal emitter, and although the detector count per distance was fixed for consistent statistics, moving the drone sampled a slightly different detector subset at each range; their unequal efficiencies could also contribute to the deviation from the expected slope (we expect efficiency variations of \SI{0.07}{\pico\watt} across the SAT1 focal plane as seen in Table 3 of \cite{Harrington:2026}). Note that for the other frequency bands, the flights would need to occur at higher altitude with the exact values being determined by the brightness of the drone with respect to saturation power, both of which increase with frequency. 

\begin{figure}
    \centering
    
    \includegraphics[width=0.465\linewidth]{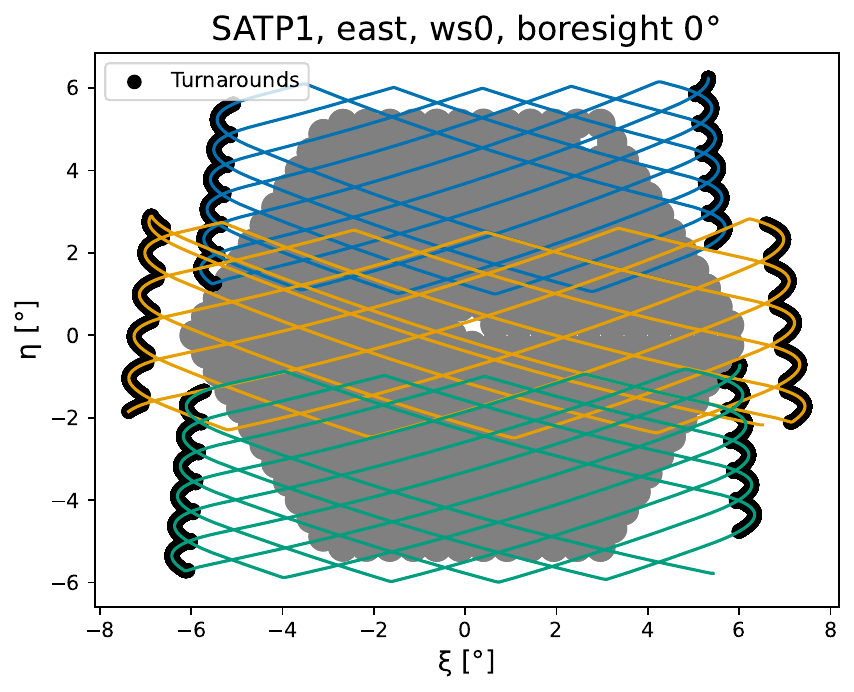}\includegraphics[width=0.465\linewidth]{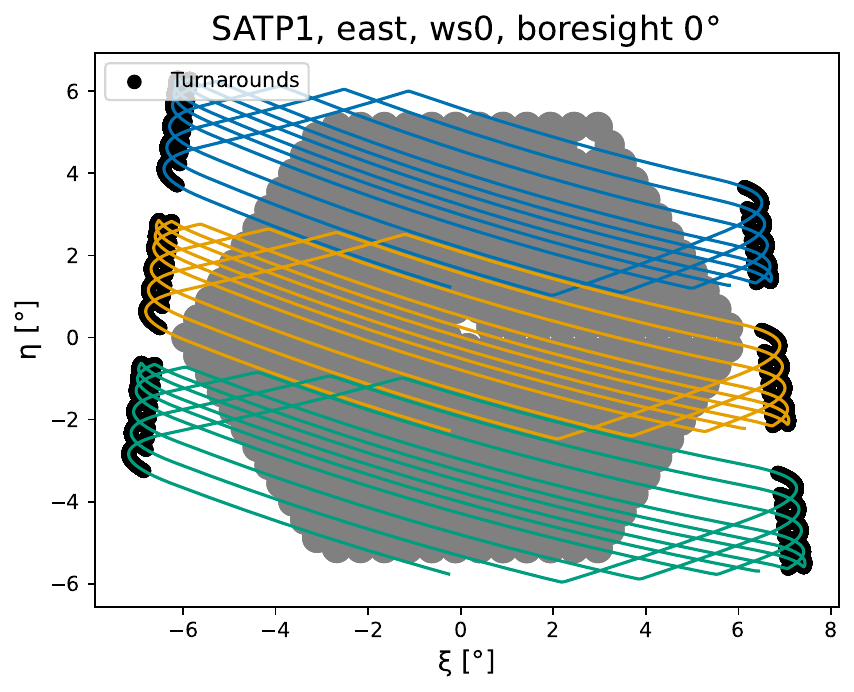}
    \includegraphics[width=0.465\linewidth]{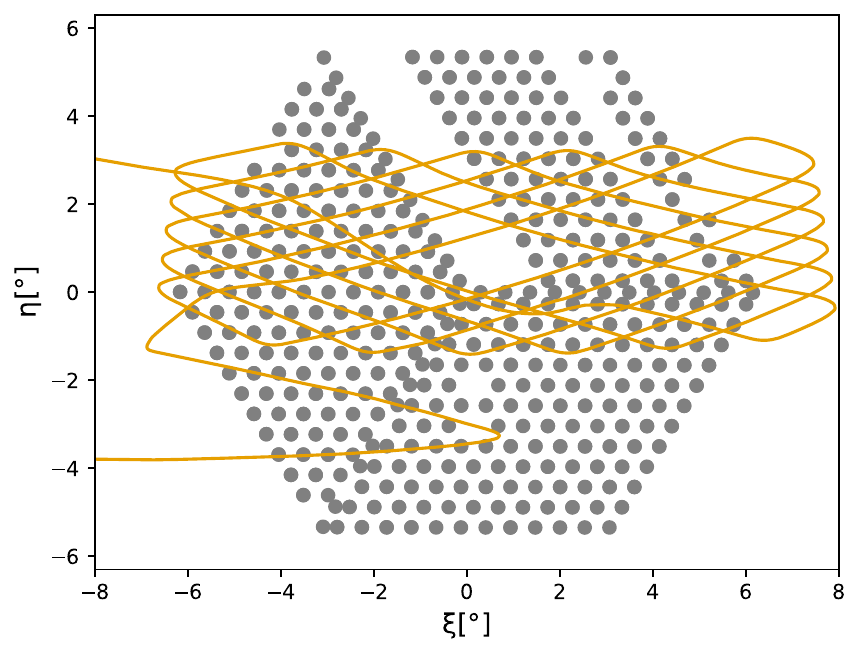}\includegraphics[width=0.465\linewidth]{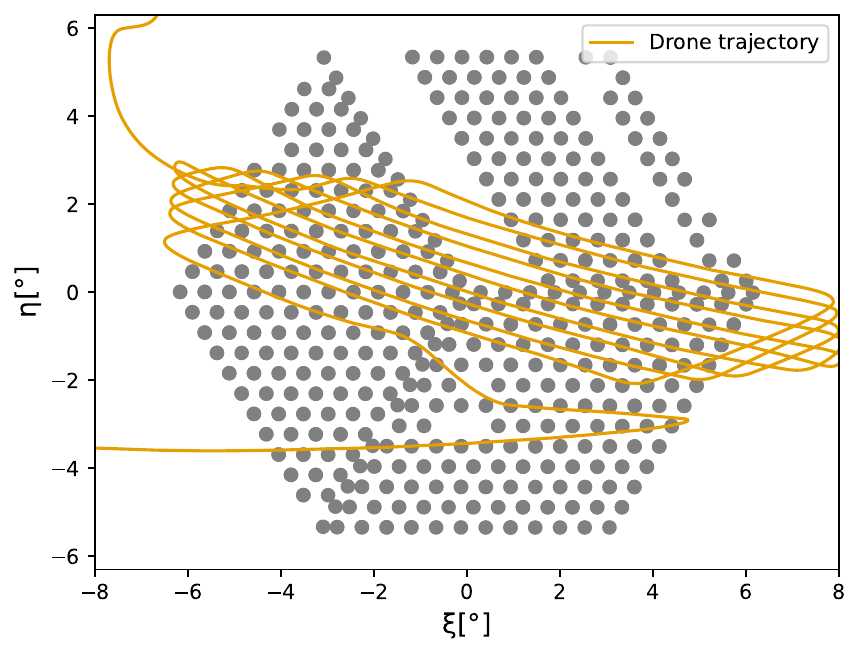}

   \caption{\label{fig:scan_patterns} Top: The expected coverage for the December 2025 flights focusing on the top, mid, and bottom (blue, orange, and green patterns) of the target wafer (ws0). The turnaround points, which are then masked in the analysis, are over-plotted with black color. Bottom: Selected examples of the achieved coverage from the December 2025 campaign. The scanning pattern of the bottom left panel also matches the achieved coverage of most flights from the April 2025 campaign.}
\end{figure}

The flight plans and scanning strategies are simulated ahead of the campaign to allow us to optimize coverage of the detectors on the SAT focal planes. An end-to-end pipeline was developed for this purpose. The pipeline produces the expected detector coverage for a given flight plan from the trajectory of the drone, the scanning motion of the SAT platforms, and the projection of the focal plane (and detector positions) onto the sky. We first generated target flight centers for the drone to fly in three positions: east (E), south (S), and west (W) of the plateau. Once in position, the drone moves along a vertical arc centered on a point equidistant between the observing telescopes. The SAT platforms scan back and forth in azimuth to achieve the grid-like movement of the drone source across the detectors at rates of \SI{0.5}{\degree/\second} and \SI{0.8}{\degree/\second} for the April and December 2025 campaigns, respectively. For the center location of the drone in each of the E/S/W flights, we calculate the required SAT boresight pointing to aim the center of each wafer at the drone, as well as the necessary span of the azimuth scan to reach each edge of the wafer. These values are unique for every wafer at each of the E/S/W drone flight centers, and are also unique to each SAT. A $\sim$ \SI{45}{\degree} `keep-out' angle from the boresight drives a sun-avoidance criteria which restricts which areas of the sky are safe to observe throughout the day. This varies between the summer and winter campaigns. Moreover, the SATs must point at boresight elevations above \SI{40}{\degree} to accommodate the safety of the spinning half wave plates and general instrument safety including cryogenics. This limits which wafers can be safely calibrated. Accessing some of these wafers will require a new drone capable of flying at an even higher elevation in the thin mountain-top atmosphere. For SAT1, we utilize boresight rotation of the telescope platform to mediate some of these concerns.

Since battery life limits each flight to less than \SI{10}{\minute} at a distance of \SI{500}{\meter}, we can not map a full wafer in a single flight. We optimize the drone and telescope schedules under the requirement of the former crossing at least 400 detectors, mapping roughly one third of a
wafer per flight. We target different parts of the wafer and different wafers depending on the scientific goals set for each campaign. Three campaigns have been conducted in total: one in December 2024 and two in April and December 2025. This paper uses only the 2025 campaign results. Both 2025 campaigns primarily targeted the center wafers as representative examples of the telescope calibration parameters. Although they are not used in the present analysis, a smaller number of measurements were also taken on selected edge wafers. The full flight statistics are presented in Appendix \ref{flight_statistics}. Figure \ref{fig:scan_patterns} shows the expected drone footprints on the focal plane together with selected examples of the achieved coverage. These plots refer to the December 2025 campaign where two different scanning patterns were tested, a denser one achieving more hits per detector and a sparser one sampling most of the detectors only once. The quality of the results was similar between these two types of scans. In the December 2025 campaign, we did \SI{93}{\giga\hertz} flights focusing also on the top and bottom parts of the center wafers, while in the April 2025 campaign only the center part of the wafers was mapped. Measurements at frequencies different than the band center were consistently performed only on the central region of the observed wafers. Note that, aside from the expanded set of scanning patterns, the system configuration remained mostly the same between the two 2025 campaigns, allowing all measurements to be analyzed consistently.

\section{Analysis}
\label{sec:analysis}

\paragraph{} Observations are processed as follows: first, we reject the ‘dark detectors’ (detectors purposefully not coupled to the sky) and any detectors sufficiently far away from the optimal bias point. The drone azimuth and elevation positions during the flight are then loaded and the time-ordered data are truncated to the time during which the drone is inside the field of view of the telescope. The positions are estimated with respect to the telescope in question using the drone GPS data, and the times during which the drone is ascending and descending are omitted. Knowing the position of the drone with respect to the telescope allows us to mask its detector crossings in the time-ordered data. After also masking the telescope turnarounds, we fit and subtract the HWP scan-synchronous signal \citep{Salatino2018, Day-Weiss:2026}. Next, the data are calibrated to picowatt (pW) \citep{Harrington:2026} and the anti-alias filter of readout and detector's time constant are deconvolved. The latter rely on the best-fit instrument parameters. With the drone data still masked, we identify and fix the step-like discontinuities in the data caused by the readout flux jumps. Some residual glitches remain within the mask, which we identify and remove. Finally, any remaining saturated detectors are identified by fitting a 2D Gaussian to the crossing peak and discarding those whose best-fit Full Width at Half Maximum (FWHM) exceeds the expected beam size by \SI{2}{\arcminute} or more, where the expected size is derived from simulations and planet data (see Table \ref{tab:beam_stats} and \cite{Day-Weiss:2026}). 

\begin{figure}
    \centering
    \includegraphics[width=0.5\linewidth]{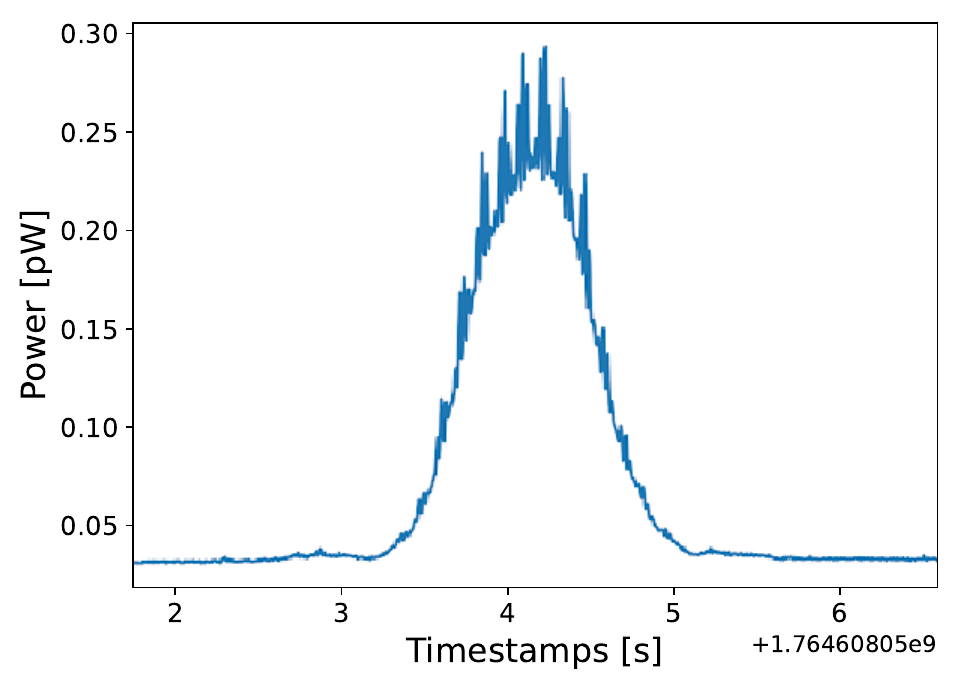}\includegraphics[width=0.5\linewidth]{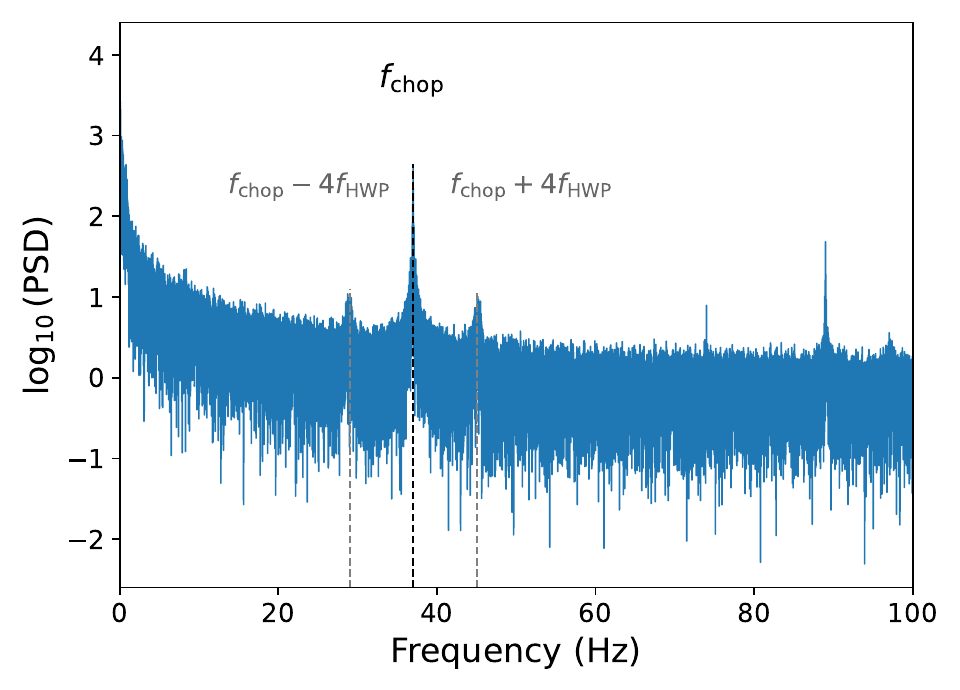}
    \caption{\label{fig:single_crossing_and_PSD}Left panel: A single drone crossing for one detector of the center wafer of SAT1 with the source emitting at \SI{93}{\giga\hertz}. The underlying shape is formed by the thermal emission of the drone and the optical beam of the telescope while the square modulation of the source signal is visible as periodically recurring bundles of spikes. Right panel: The PSD of all stacked crossings of all detectors for the same flight. Note the large peak at the modulation frequency and the symmetric peaks around it at $\pm4f_{\mathrm{HWP}}$.}
\end{figure}

A typical drone crossing is captured on the left panel of Fig. \ref{fig:single_crossing_and_PSD}, showing a \SI{5}{\second} sample of one detector from a SAT1 observation with the source emitting at \SI{93}{\giga\hertz}. The underlying Gaussian-looking shape corresponds to the thermal emission of the drone while the bundles of spikes are the result of the source being `on'. The source signal is modulated by a square wave at $f_{\mathrm{chop}}=37.00015$ \SI{}{\hertz}, so the source contributes to the data only during the wave's positive half-cycles. The HWP further modulates the amplitude of this signal. The strength of the source emission is also shown in the right panel of the same figure through the power spectral density (PSD) of all stacked crossings of all detectors during that same flight. A large peak is clearly visible at $f_\mathrm{chop}$ and the smaller-amplitude, symmetric modulation peaks due to the HWP at $\pm4f_{\mathrm{HWP}}$, where $f_{\mathrm{HWP}}$ is the HWP rotation frequency. 

To reconstruct the true polarization of the source, one needs to double-demodulate the data. However, for temperature studies it is sufficient to demodulate the chopped signal and select frequency components lower than the HWP spin rate. We process as follows: (i) The stacked signal from all crossings of all detectors is multiplied by $\exp (2j \pi f_{\mathrm{chop}}t)$, (ii) The resulting signal is low-passed to a cut-off equal to $1.9*f_{\mathrm{HWP}}$, and (iii) The phase is defined as the signal-weighted average of the arctangent of the imaginary and the real part of the complex baseband signal. Initially, we assume perfect knowledge of the chopping frequency, but we then fit for it by minimizing the phase slope as a function of time. The fitted frequency is also cross-checked with the one recovered from the source data. Finally, we employ the best-fit chopping frequency and phase to demodulate the original signal and apply a low-pass filter. Note that the source signal occupies only a narrow band, $\sim 10^{-5}$ \SI{}{\hertz} wide, around the chopping frequency; demodulating at a frequency offset by more than this bandwidth therefore loses the signal. Since the low-pass filter allows a much broader frequency range, some leakage from other temperature components is expected. However, after accounting for detector efficiency, the expected and measured source power agree to within $5\%$ across all observations (see \cite{Bizzarri_2026prep}). The total filtering bias on the reconstructed beam is estimated via simulations produced using an updated version of the code described in \cite{Coppi:2025}. The simulations include the real scan strategy, realistic detector and atmospheric noise, HWP modulation, and the monochromatic source signal modulated at $f_{\mathrm{chop}}$. When applied to simulations, the pipeline yields an unbiased reconstruction of the main lobe and reproduces the beam wings to within $0.5\%$ down to at least \SI{-30}{\decibel}.
    
After demodulating the data, we bin them into maps using a version of the planet map-maker described in \cite{Hasselfield_2013, Lungu_2022, Dachlythra_2024} that includes the drone pointing and corrects for the pointing deflection that arises from the HWP that was discovered by drone studies \citep{Dunner_2021, Nati_2017, Harrington:2026}. Specifically, the beam maps are reconstructed in a drone-centered tangent-plane projection whose axes are the native focal-plane coordinates $(\xi, \eta)$, where $\xi$ is parallel to increasing azimuth and $\eta$ parallel to increasing elevation as seen along the telescope boresight. Each detector is placed by its offset $(\xi, \eta)$ from the boresight, and the boresight pointing is rotated into the frame of the drone so that the source lies at the origin. A key difference is that, in the drone case, no Principal Component Analysis (PCA) filtering of the atmosphere is needed. This is due to the source signal being modulated at a specific frequency. In other words, the modulation is what drives the high SNR of the measurements and not the high power of the source itself. The latter is in fact set to a value about twice the peak power measured from Jupiter maps \citep{Day-Weiss:2026}. 

\begin{figure}
    \centering
    \includegraphics[width=0.5\linewidth]{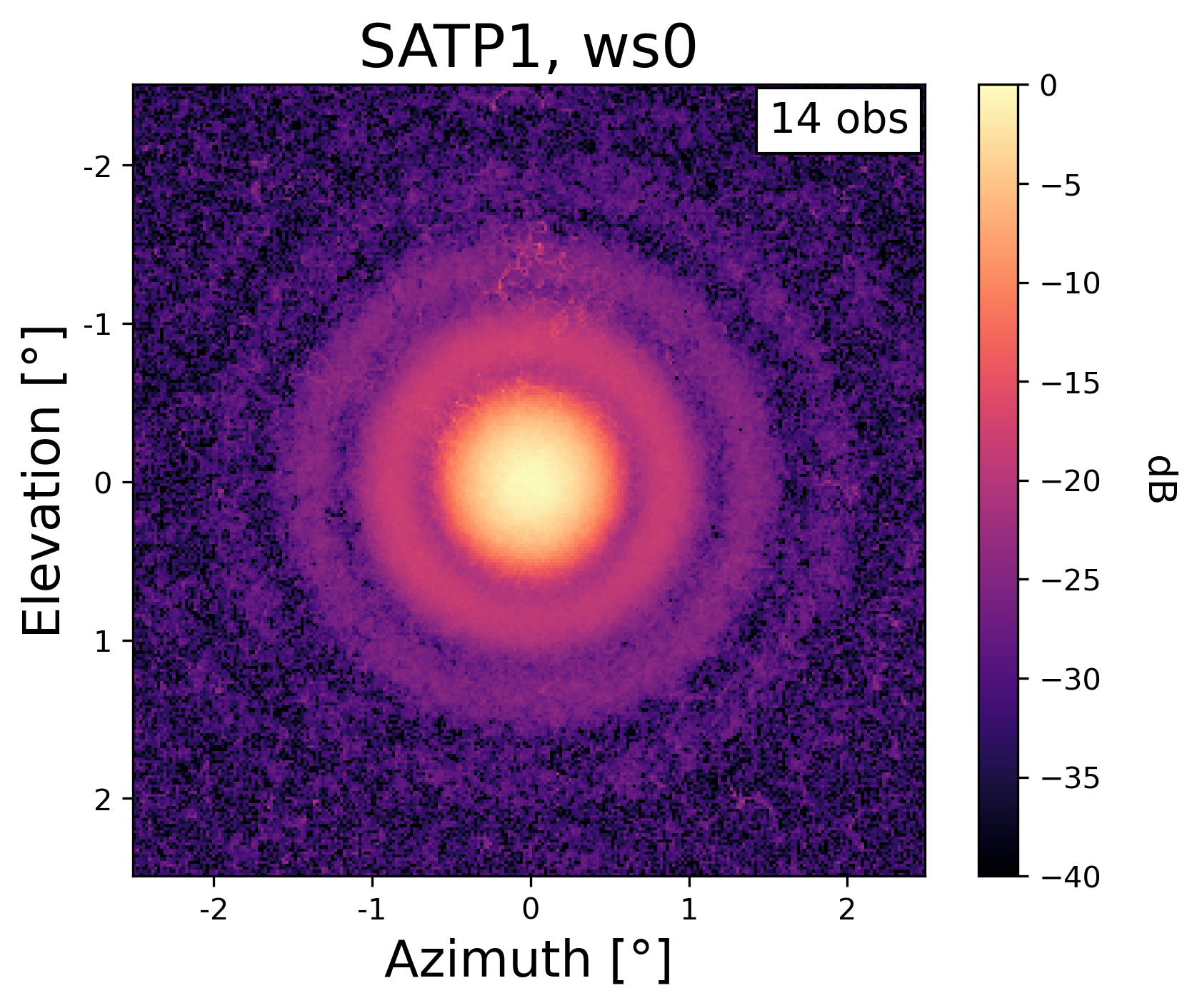}\includegraphics[width=0.5\linewidth]{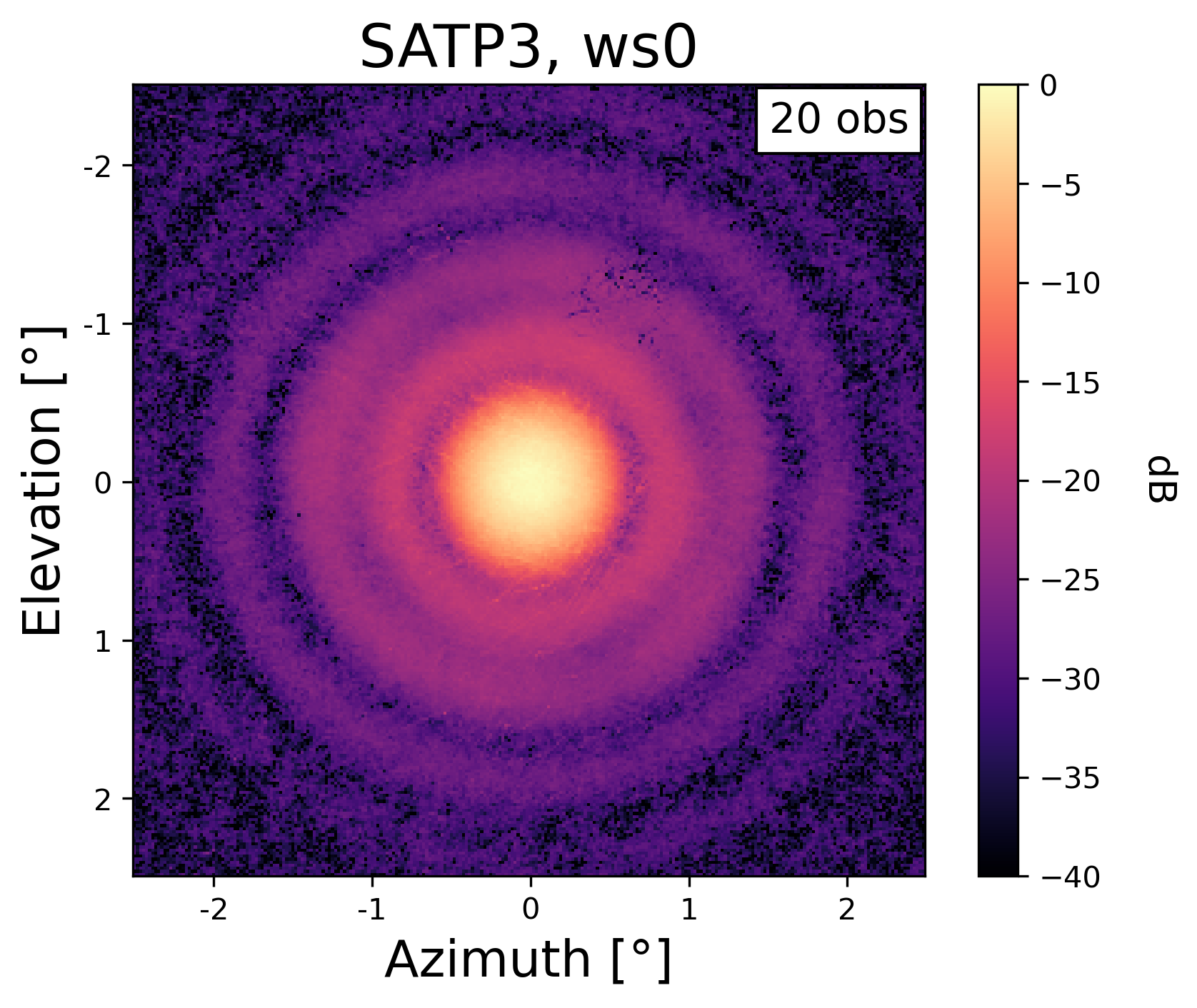}
    \caption{\label{fig:93_ghz_maps}Inverse-variance weighted maps from all observations at \SI{93}{\giga\hertz} of the two 2025 campaigns for SAT1 (left) and SAT3  (right). Note the slightly different structure of the measured lobes between the beams of the two telescopes.}\end{figure}

Representative examples of reconstructed temperature maps are shown in Figure \ref{fig:93_ghz_maps}. These correspond to inverse-variance weighted \SI{93}{\giga\hertz} observations from the central wafers of SAT1 and SAT3, combining data from both 2025 campaigns. Maps at all measured frequencies are shown in Appendix \ref{single_f_beam_maps}. We focus on the middle parts of the central wafers (ws0) of the SAT1 and SAT3  telescopes, as these regions of the focal plane were the most densely populated with multi-frequency observations across both campaigns. All results of Section \ref{sec:results} are presented through the beam radial profiles and the corresponding harmonic transforms, except for the measured beam size and ellipticity that are fitted from the maps.

\section{Results}
\label{sec:results}

\subsection{Beam fitting pipeline}
\label{sec:beam_fitting_pipeline}

The beam profiles are constructed by radially binning the maps after fitting a 2D Gaussian to define the map centers. They are also normalized to 1 at $\theta=0$. Each profile is then fitted to a model composed of a Gaussian function for the main lobe, $B_{\mathrm{core}}(\theta)$, a combination of a decaying envelope (generalized Moffat profile \citep{moffat_1969}), $E(\theta)$, and a damped oscillation model, $R(\theta)$, for the signal dominated region of the lobes, $B_{\mathrm{lobes}}(\theta)$, and a modified $1/\theta^{3}$ function for the wing, $B_{\mathrm{tail}}(\theta$). The piece-wise profile model is defined as following:

\begin{equation}
B_{\mathrm{core}}(\theta) = \exp\left(-\frac{\theta^2}{2\sigma_\mathrm{g}^2}\right),
\end{equation}

\begin{equation}
\left.
\begin{aligned}
E(\theta) &= \frac{A}{\left(1 + (\theta/\theta_s)^p\right)^q} + B \\
R(\theta) &= \left( 1 + c e^{-\beta \theta} J_0(\kappa\theta + \phi) \right)^2
\end{aligned}
\right\} \hspace{0.3cm} B_{\mathrm{lobes}}(\theta) = E(\theta)R(\theta),
\end{equation}
\begin{equation}
B_{\text{tail}}(\theta) = B(\theta_2)\left(\frac{\theta_2 + \theta_0}{\theta + \theta_0}\right)^3.
\end{equation}
\\

\begin{figure}[h]
    \centering
    \includegraphics[width=0.74\linewidth]{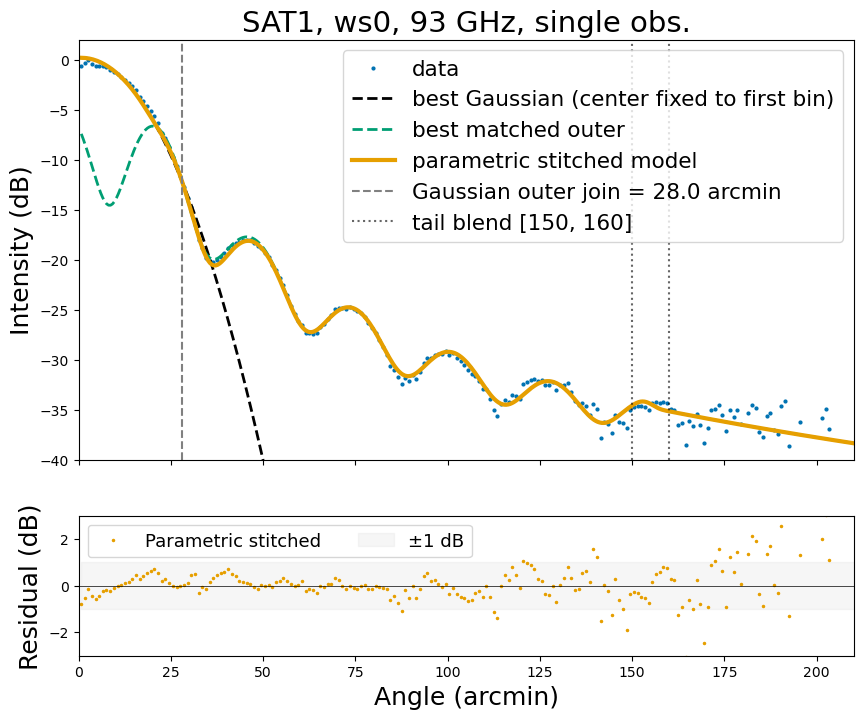}
    \includegraphics[width=0.74\linewidth]{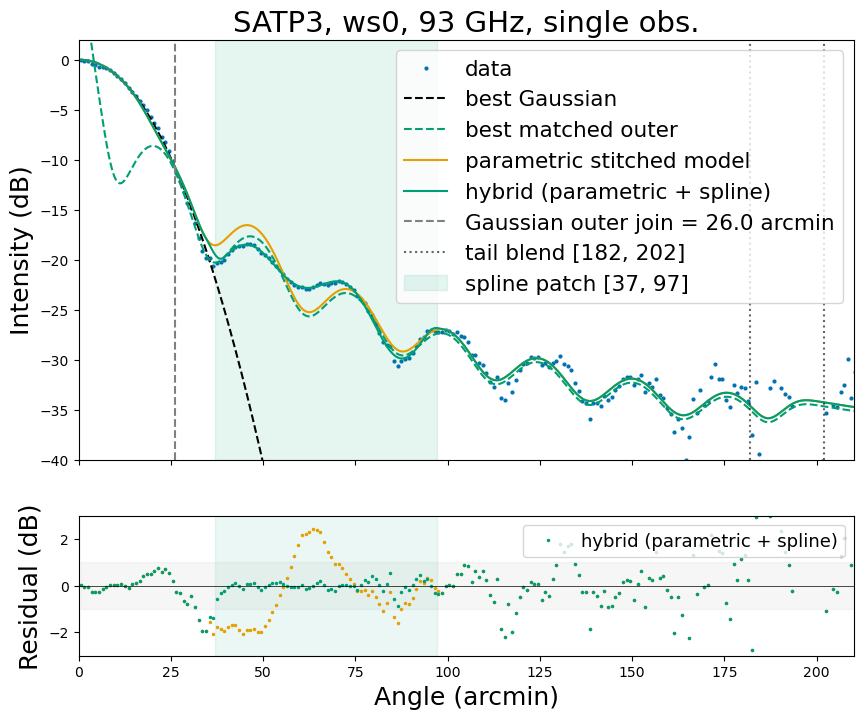}
    \caption{\label{fig:fit_profiles_SAT1_SAT3} Fitted profiles for two flights targeting the center wafers of SAT1 (top) and SAT3  (bottom) with the drone source emitting at \SI{93}{\giga\hertz}. The parametric model (orange lines) adapts well to the SAT1 data but produces a prominent residual feature over the first two lobes of the SAT3  example data. A hybrid model (green solid line) applying a spline fit only over that particular angular range accounts for the residual bias. }
\end{figure}

The parameter $\sigma_{g}$ naturally refers to the standard deviation of the
Gaussian distribution that assumes the first profile bin as the center. The
parameter $\theta_{s}$ in the decaying part of the sidelobe model sets the scale
angle where the decay begins to noticeably roll over, while $p$ and $q$ control
how sharply the denominator turns on and the overall steepness of the decay once
roll-off has started. An additional parameter $A$ sets the amplitude, while $B$
captures the baseline noise, which is almost consistent with zero. The
oscillatory sidelobe structure of a diffraction-limited beam is described by the
Airy pattern, which is built on the first-order Bessel function $J_{1}$ and fixes
the ring positions to those of an ideal circular aperture. Since our aim is to characterize deviations from this ideal pattern 
we instead adopt a phenomenological damped oscillation based on the
zeroth-order Bessel function of the first kind, $J_{0}$, with a free spatial frequency, $\kappa$, and phase, $\phi$. 
The $J_{0}$ term is further modified by the oscillation amplitude and damping rate, $c$ and $\beta$, respectively. For the beam wing, we default to the standard $1/\theta^{3}$ expectation
\citep{Lungu_2022, Dachlythra_2024} but add a positive shift in angle,
$\theta_0$, that softens the tail near the anchor angle, $\theta_2$, from which we apply the wing model. The sidelobe and tail models are stitched using a smooth cubic blend:

\begin{equation}
w_1(x) = 1 - \left(3x^2 - 2x^3\right),
\qquad
x = \frac{\theta - \theta_1}{\theta_2 - \theta_1},
\
\end{equation}

\noindent where $\theta_{1}$ and $\theta_{2}$ define the lower and upper limit of the blend region. The two models are synthesized as:

\begin{equation}
B_{\text{outer}}(\theta)=w_1(\theta)\,B_{\text{sidelobe}}(\theta)+\left(1-w_1(\theta)\right)B_{\text{tail}}(\theta).
\end{equation}

We finally match the outer core and tail model to the main beam Gaussian by matching amplitude and log-slope using a window of width
W on each side of a reference angle $\thj$ (kept below one ripple period so the match remains local and avoids the Bessel oscillation). With

\begin{equation}
  s = \frac{B_{\mathrm{core}}(\thj)}{B_{\mathrm{outer}}(\thj)},
  \qquad
  \alpha = d_{\mathrm{in}} - d_{\mathrm{out}},
\end{equation}
where $d_{\mathrm{in}}, d_{\mathrm{out}}$ are the least-squares slopes of $\ln B_{\mathrm{core}}$ and $\ln(sB_{\mathrm{outer}})$ on $[\thj - W, \thj]$ and $[\thj, \thj + W]$, the matched outer model is
\begin{equation}
  \Bouterm(\theta) = s\,\Bouter(\theta)\,\exp\!\bigl[\alpha\,(\theta - \thj)\,D(\theta)\bigr],
  \qquad
  D(\theta) = \exp\!\left[-\bigl((\theta - \thj)/\Delta\bigr)^{2}\right] .
\end{equation}
This enforces $\Bouterm(\thj) = \Bcore(\thj)$ and $(d\ln \Bouterm / d\theta)|_{\thj} = d_{\mathrm{in}}$, with the slope correction damped to zero on scale $\Delta$ so the tail is unaffected. The final model is the cubic smoothstep blend over $[\tha, \thb] = [\thj - \Delta, \thj + \Delta]$:
\begin{equation}
  w_{2}(\theta) = 1 - \bigl(3y^{2} - 2y^{3}\bigr),
  \qquad
  y = \frac{\theta - \tha}{\thb - \tha},
\end{equation}
\begin{equation}
  \Bfinal(\theta) = w_{2}(\theta)\,\Bgauss(\theta) + \bigl(1 - w_{2}(\theta)\bigr)\,\Bouterm(\theta) .
\end{equation}
\\

\par To select the optimal parameter values, we use the Bayesian Information Criterion (BIC) on the log-intensity residuals \footnote{For comparison, the Akaike Information Criterion would penalize complexity by $2k$ rather than $k\ln n$:
\begin{equation}
  \mathrm{AIC} \;=\; n\,\ln\!\left(\frac{\RSSlog}{n}\right) \;+\; 2k .
\end{equation}
For sample sizes in the hundreds, $\ln n \gtrsim 5$, making BIC considerably more conservative than AIC --- the appropriate bias when the goal is a parsimonious semi-empirical model.} . With the residual sum of squares in log-intensity constructed as
\begin{equation}
  \RSSlog \;=\; \sum_{i=1}^{n}\bigl[\ln B_{\mathrm{raw}}(\theta_{i}) - \ln B_{\mathrm{final}}(\theta_{i})\bigr]^{2} ,
\end{equation}
we employ the BIC criterion (Gaussian-error log-residual approximation)
\begin{equation}
  \mathrm{BIC} \;=\; n\,\ln\!\left(\frac{\RSSlog}{n}\right) \;+\; k\,\ln n 
\end{equation}
with $k = 10$ effective parameters: one Gaussian, $\sigma_{g}$, and nine outer $\{A, \ths, p, q, B, c, \beta, \kappa, \phi\}$. The joint angles, at which the model transitions between its different components, are determined by a coarse grid search rather than being treated as free parameters. Reasonable changes in the joint angles do not affect the resulting beam transfer function. Specifically, the joint angle between the Gaussian and the near-lobe model varies only slightly across the different observations, while a large change in the second joint angle — where the noise-dominated region begins — affects only features well below \SI{-30}{\decibel}. Note that $\theta_{0}$ is a fixed regularization constant in the tail, softening the $1/\theta^{3}$ term so it can not blow up; it is part of the functional form, not a degree of freedom the optimizer explores. 

Figure \ref{fig:fit_profiles_SAT1_SAT3} shows examples of beam radial profiles for two flights for the center wafers of SAT1 (top) and SAT3  (bottom) with the source emitting at \SI{93}{\giga\hertz}. The parametric fitted model is overplotted in orange in both cases. While the described model adapts well to the SAT1 data, it leaves a pronounced $\sim$2 dB residual over the first two sidelobes of the SAT3 data, between \SI{40}{\arcminute} and \SI{100}{\arcminute} (bounded by the first and third beam minima). The amplitude and centre of this feature scale with frequency, as shown in Appendix \ref{SAT3_sl_bump_vs_f}. Faced with this residual we have two choices: extend the parametric model with an additional component that captures the residual and add it back, or fit the measured profile directly over this range with a spline. Since, as discussed in Section \ref{sec:single_f_beams}, we expect this residual structure to peak at a frequency outside the band we are mapping, we consider the second option the appropriate one and fit the data with a spline over the first two lobes. The minima are located as peaks of the negated, Gaussian-smoothed ($\approx$\SI{3}{\arcminute}) log-profile, using a prominence threshold of \SI{0.5}{\decibel} and a minimum spacing of \SI{10}{\arcminute}, so that noise wiggles are not mistaken for beam lobes. We then fit a spline of degree $k=4$ to the log-profile (in dB) between the first and third minima, using \texttt{UnivariateSpline} with smoothing factor $s=0.05$. No blending with the parametric model is required in this case, since the spline coincides with it at the two bounding minima. The resulting hybrid model (green) accounts for the residual over the shaded angular range, which corresponds to the extent of the spline fit.

\subsection{Drone-calibrated single frequency beams}
\label{sec:single_f_beams}

\begin{figure}
    \centering
    \includegraphics[width=0.5\linewidth]{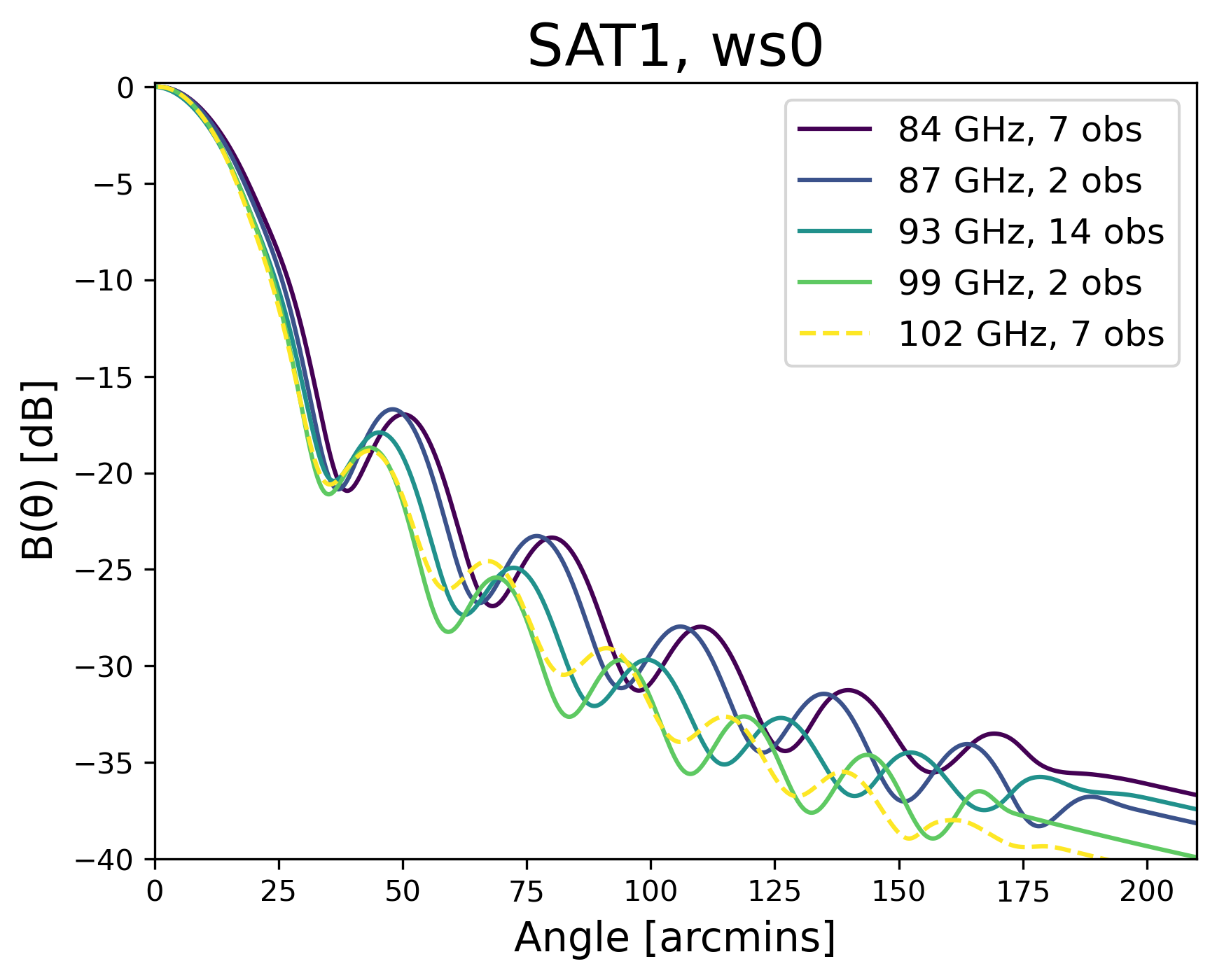}
    \centering
    \includegraphics[width=0.49\linewidth]{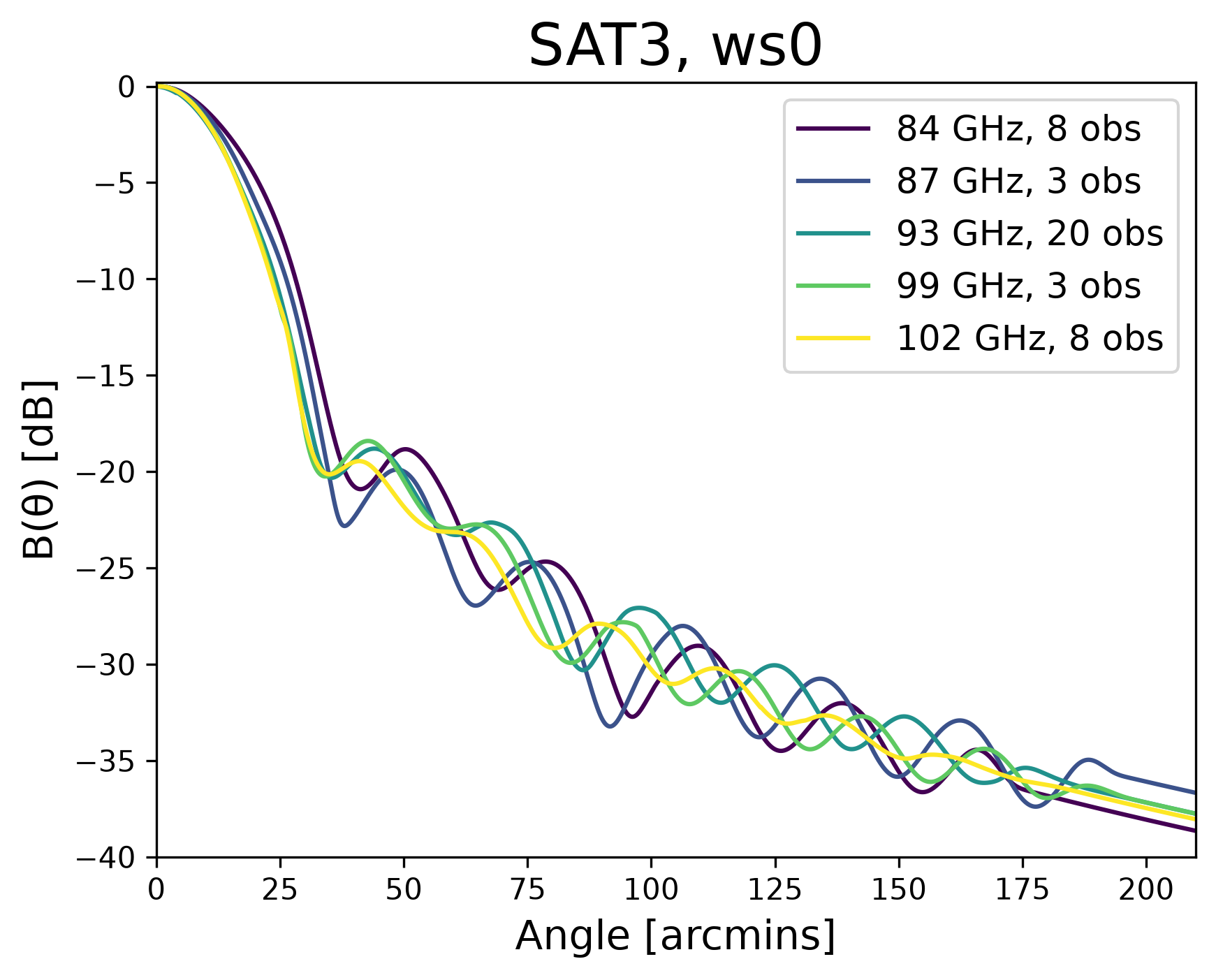}

\caption{\label{fig:radial_profiles_SAT1_and_SAT3} Left: Fitted beam profiles using the parametric model described in Section \ref{sec:beam_fitting_pipeline} for all observations targeting the middle part of the center wafer of SAT1. Note the \SI{102}{\giga\hertz} beam (yellow dashed curve) deviating visibly from the frequency scaling between the other profiles. Right: Fitted beam profiles for SAT3 using the same parametric model, plus a spline correction applied roughly over the angular range of the second and third lobes.}

\end{figure}

This section presents the fitted beam profiles and corresponding beam transfer functions as a function of frequency for the middle part of the SAT1 and SAT3 center wafers. For SAT1, the beam profiles exhibit the expected frequency scaling, with the overall beam narrowing systematically toward higher frequencies as seen in the left panel of Figure \ref{fig:radial_profiles_SAT1_and_SAT3}. A similar trend is observed for the SAT3 core; however, the near lobe structure displays a more complex frequency dependence and does not scale as simply (see right panel of the same figure). At \SI{84}{\giga\hertz}, the first two lobes remain clearly distinct, although their separation becomes less pronounced as the frequency increases. By approximately \SI{144}{\giga\hertz}, these lobes will merge into a single prominent feature, as can already be established from the band-averaged Jupiter profiles shown in \cite{Harrington:2026} and will be further discussed in \cite{Day-Weiss:2026}. This non-trivial scaling justifies our choice to use spline fitting over the affected angular range. An exception to the beam frequency scaling for SAT1 is observed at \SI{102}{\giga\hertz}, where the beam appears significantly noisier than \SI{99}{\giga\hertz}, obscuring much of its structure. Since the source power was approximately consistent across frequencies, and no substantially worse weather was present, the discrepancy is unlikely to be atmospheric or source-driven. The most plausible remaining candidate is enhanced standing waves from the polarizing wire grid at that particular frequency; we cannot fully exclude leakage of the drone's thermal body emission, though its smooth increase with frequency makes it a poor match for a feature localized to a single channel. Note that both figures display the inverse-variance-weighted average profiles per frequency.

The average beam size and ellipticity per telescope and frequency are estimated before constructing the profiles, by fitting a 2D elliptical Gaussian to the beam maps. Table \ref{tab:beam_stats} presents the estimated beam properties along with their uncertainties calculated as the standard deviation across all flights per measured frequency. We also include the expected best-fit values from simulations at the same frequencies. The Physical Optics (PO) simulations are produced with \texttt{TICRA TOOLS} assuming a 3-lens refracting telescope and a pixel \SI{3}{\centi\meter} away from the center of the focal plane \citep{Dachlythra_2024}, corresponding to half the radius of ws0. From the table we conclude that the measured beams scale roughly as expected and show no significant ellipticity. We quantify the ellipticity as
\begin{equation}
\varepsilon = \frac{\max(\sigma_{x},\sigma_{y}) - \min(\sigma_{x},\sigma_{y})}{\sigma_{x}+\sigma_{y}},
\end{equation}
where $\sigma_{x}$ and $\sigma_{y}$ are the fitted Gaussian widths along the two beam axes\footnote{We adopt this core-width (Gaussian-$\sigma$) definition of beam ellipticity because it gives a direct, intuitive characterization of the beam core, which is the part most relevant for optical performance and for comparison with diffraction-limited expectations.}. This is identical to the definition of beam ellipticity employed in \cite{Harrington:2026}. Both telescopes show a shallower FWHM–frequency scaling than expected: the measured FWHM is below the prediction up to $\sim$ \SI{93}{\giga\hertz} and above it at higher frequencies. The deviation is more pronounced for SAT1. However, more robust statistics of the band-averaged beam properties, incorporating a substantially larger number of measurements across the full focal plane, will be presented by \cite{Day-Weiss:2026}. 

\begin{table}[h!]
    \centering
    
    \begin{tabular}{|c|c|c|c|c|}
    \hline
       Telescope, wafer, Freq (\SI{}{\giga\hertz}) & FWHM & FWHM$_{\mathrm{sim}}$ & $\varepsilon$ & $\varepsilon_{\mathrm{sim}}$\\

       \hline
       SAT1, ws0, 84 & 29.7 $\pm$ 0.4 & 30.7 & 0.017 $\pm$ 0.009 & 0.030\\

       \hline
       SAT3, ws0, 84 &  30.5 $\pm$ 0.2 &  &  0.014 $\pm$ 0.005 & \\

       \hline
       SAT1, ws0, 87 & 28.4 $\pm$ 0.1 & 29.7 & 0.006 $\pm$ 0.002 & 0.030  \\

       \hline
       SAT3, ws0, 87 & 29.2 $\pm$ 0.1 & & 0.008 $\pm$ 0.005 & \\

       \hline
       SAT1, ws0, 93 & 27.9 $\pm$ 0.4 & 27.6 & 0.017 $\pm$ 0.009 & 0.03 \\

       \hline
       SAT3, ws0, 93 & 27.5 $\pm$ 0.6 & & 0.017 $\pm$ 0.009 & \\

       \hline
       SAT1, ws0, 99 &  26.7 $\pm$ 0.3 & 26.1 & 0.014 $\pm$ 0.000 & 0.03  \\

       \hline
       SAT3, ws0, 99 & 26.2 $\pm$ 0.3 & & 0.022 $\pm$ 0.011 & \\

       \hline
       SAT1, ws0, 102 & 25.9 $\pm$ 0.4 & 25.3 & 0.024 $\pm$ 0.001 & 0.03\\

       \hline
       SAT3, ws0, 102 & 26.0 $\pm$ 0.5 & & 0.027 $\pm$ 0.011 & \\
       
       \hline
       \end{tabular}
       \caption{\label{tab:beam_stats} The average beam size (FWHM) and ellipticity ($\varepsilon$) with their associated uncertainty as estimated by applying a 2D elliptical Gaussian on beam maps from both 2025 campaigns. These are presented per telescope and frequency, and compared against the expected values ($\mathrm{FWHM}_{\mathrm{sim}}$, $\varepsilon_{\mathrm{sim}}$) from simulations at the same frequencies.}
       
\end{table}

\begin{figure}

    \centering
    \includegraphics[width=0.49\linewidth]{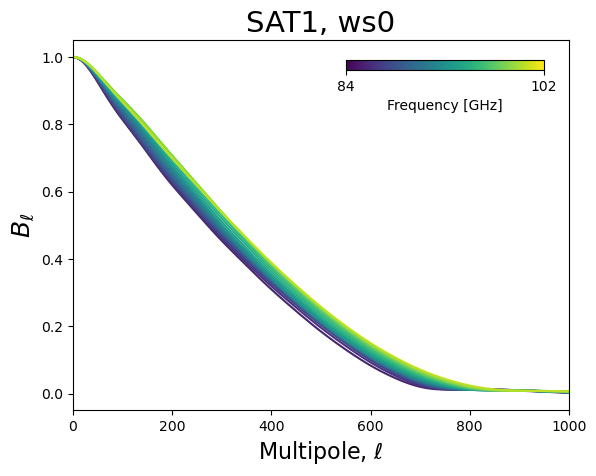}
     \includegraphics[width=0.49\linewidth]{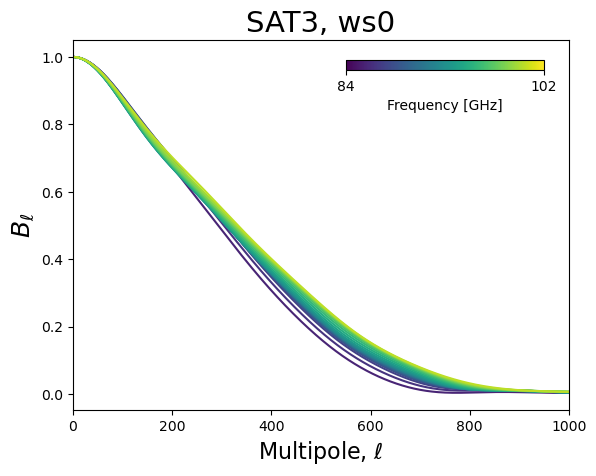}
     \caption{\label{fig:bells_continuous_SAT1_and_SAT3} The SAT1 ws0 (left) and SAT3 ws0 (right) continuous beam transfer functions between \SI{84}{} (dark blue) and \SI{102}{\giga\hertz} (bright yellow) interpolated to the instrument's resolution. Note the slope of the SAT3 beam transfer functions decreasing as the frequency increases.}
     
\end{figure}

We use the harmonic transforms of the average fitted beam profiles at each frequency for SAT1 ws0 and SAT3 ws0 to construct a continuous beam transfer function between the lowest and highest measured frequencies. To do this, we interpolate the harmonic transforms of the fitted beam profiles to the instrument resolution using splines. While this spline-based approach can overfit the more complex beam profiles and partially model noise, it works well for the monotonic, smoothly varying transfer functions considered here. Figure \ref{fig:bells_continuous_SAT1_and_SAT3} shows the continuous beam transforms from the lowest to the highest measured frequencies at the instrument’s resolution for SAT1 ws0 (left) and SAT3 ws0 (right). For SAT1 ws0, the nearly trivial frequency scaling suggests that the cosmological spectra can be corrected using only the band-averaged beam. In contrast, SAT3 is expected to require an additional parametrized chromatic correction to capture changes in the slope of the harmonic transform over multipoles $\ell \approx 50-300$. The departure of the SAT3 beam from the expected scaling is frequency-dependent, with the lower-frequency harmonic transforms falling off noticeably faster with $\ell$. As mentioned in Section~\ref{sec:system_overview}, the main difference in optical design between the two telescopes is the AR coating applied to the HWP alumina and to the 4\,K alumina IR filter. We therefore consider the coating the most likely cause of this departure, although a contribution from forebaffle reflections cannot be excluded. Because this beam feature is nevertheless well constrained by the measurements, it can be properly accounted for in the analysis and is therefore not expected to affect the science results.

Normalizing to $\ell \sim 50$ and employing multipoles between 50 and 400 allows us to test the standard frequency scaling of the beam FWHM:

\begin{equation}
\label{eq:FWHM_scaling}
  b_{\ell,\nu} = \exp\!\left[
      -\tfrac{1}{2}\,\ell(\ell+1)
      \left(
        \frac{\mathrm{FWHM}(\nu_0)\left(\dfrac{\nu}{\nu_0}\right)^{-\alpha/2}}
             {2\sqrt{2\ln 2}}
      \right)^{\!2}
    \right], \hspace{0.25cm} \nu_{0} = 93 \hspace{0.05cm} \mathrm{GHz}.
\end{equation}

\noindent This beam chromaticity model was also adopted in \cite{Giardello_2024} for a high-resolution instrument following the design of the SO Large Aperture Telescope (LAT). Applying it to our measurements, for SAT1 we find $\alpha = 1.85 \pm 0.09$, consistent with nearly diffraction-limited optical performance ($\alpha = 2$). A fit of comparable quality is not currently achievable for SAT3 over the same multipole range, but restricting the fit to $\ell \gtrsim 150$ yields $\alpha = 1.35 \pm 0.10$. Clearly, a more informed fitting model is required for the more complex beams of the SO SATs.

\subsection{Comparison with planet measurements and simulations}
\label{sec:planet_comparison}

Although there are many advantages to frequency-resolved instrument calibration, describing the band-averaged beam this way can be challenging as it requires sufficient frequency coverage. The five measured frequencies used throughout this analysis were chosen in this way to uniformly span a $20\%$ relative-bandwidth frequency range around the center frequency of the MF1 band of the SO SATs ($\sim$ \SI{93}{\giga\hertz}). Preliminary estimates of the telescopes' \SI{90}{\giga\hertz} bandpass \citep{Harrington:2026} suggest all five frequencies correspond to high transmission values; however, that is not necessarily a sufficient criterion and there could be other factors to consider, such as the shape of the bandpass. A sloped bandpass, for instance, breaks the symmetry of the sampling about the band center. A more structured bandpass could demand measurements wherever its slope changes to facilitate interpolation and provide a continuous description of the beam and gain across the band. At present, the most reliable way to determine both the number and the selection of frequency measurements needed to approximate the broadband calibration is by comparison with Jupiter observations from the SATs. Preliminary Jupiter beam analysis has already been presented in \cite{Harrington:2026} while \citep{Day-Weiss:2026} will expand on the more intricate beam features and show azimuthally averaged profiles that extend to larger angles. The latter are the ones used for this comparison together with optical simulations that help guide our modeling choices. Note that the simulated beams should not be regarded as ground truth either, as the optical model may be too simplified to capture all the details of the true beams; they mainly serve to indicate how much more oscillatory structure is expected when the beam is built from only a subset of frequencies.

We present the comparison between the planet (real band-averaged) and drone (synthetic band-averaged) beams in terms of both their radial profiles and transfer functions. For the drone data, each monochromatic beam, $B(\theta, \nu)$, is first normalized to unity at the beam center, thereby removing the effect of source-power variations between flights. The preliminary bandpasses, $\tau(\nu)$, derived from on-site FTS measurements of the LAT and adjusted for the known differences in the SAT design \citep{Harrington:2026}, are then applied as frequency-dependent weights. The resulting beams are averaged as described below:

\begin{equation}
\label{chrom_beam}
B(\theta) = \frac{\int \tau(\nu)B(\theta, \nu)d\nu}{\int \tau(\nu) d\nu},  \hspace{0.9cm} \nu = 84, 87, 93, 99, 102 \hspace{0.1 cm} \mathrm {GHz}.
\end{equation}

\noindent The caveat to this approach is the significant dependence on the bandpass assumption, which is why in the future only the expected source power will be removed before averaging across the different frequencies. This, however, requires accurate estimates of the end-to-end detector efficiency per frequency and represents an ongoing effort \citep{Bizzarri_2026prep}. For the planet data,  the profiles are not fitted with the same parametric model as the drone; instead Bessel functions are employed up to an optimal transition radius then the $1/\theta^{3}$ fit takes over \citep{Day-Weiss:2026}. The planet and drone beams are also compared against the band-averaged radial profile from PO simulations. Far-field beam maps are first simulated for a representative frequency sample and then interpolated either to the true frequency resolution of the SO SATs or the five frequencies measured with the drone. Averaging these two sets of interpolated beams after weighting as shown in Equation \ref{chrom_beam} yields the simulated band-averaged beams to be compared with the planet and drone cases. Note that this comparison will only use the average drone and planet products as it is meant to be a demonstration of the feasibility of the frequency-resolved calibration scheme, rather than the analysis from which the precise beam covariance and instrument requirements will be set.

\begin{figure} [H]
    \centering
    \includegraphics[width=0.7\linewidth]{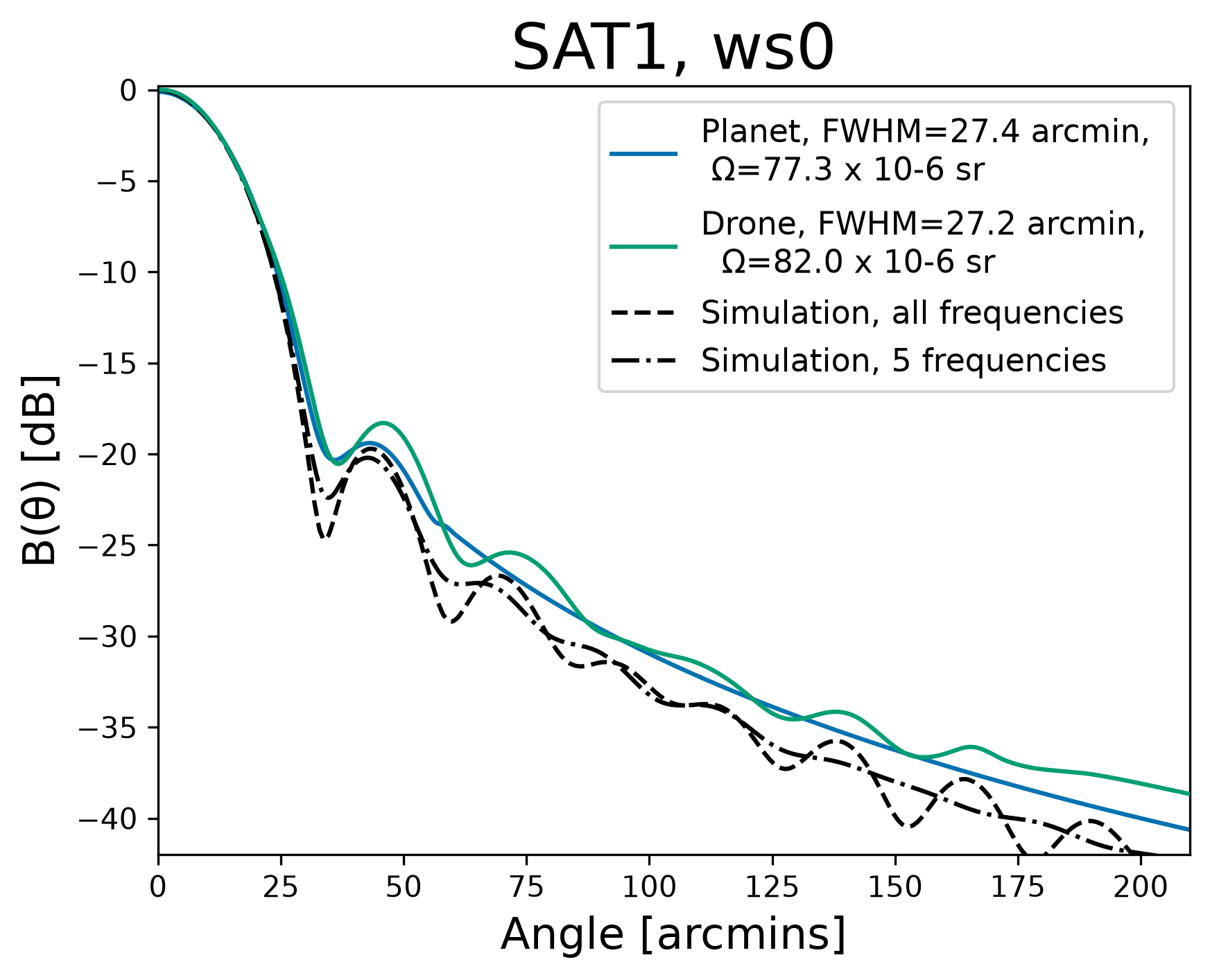}
    \centering
    \includegraphics[width=0.7\linewidth]{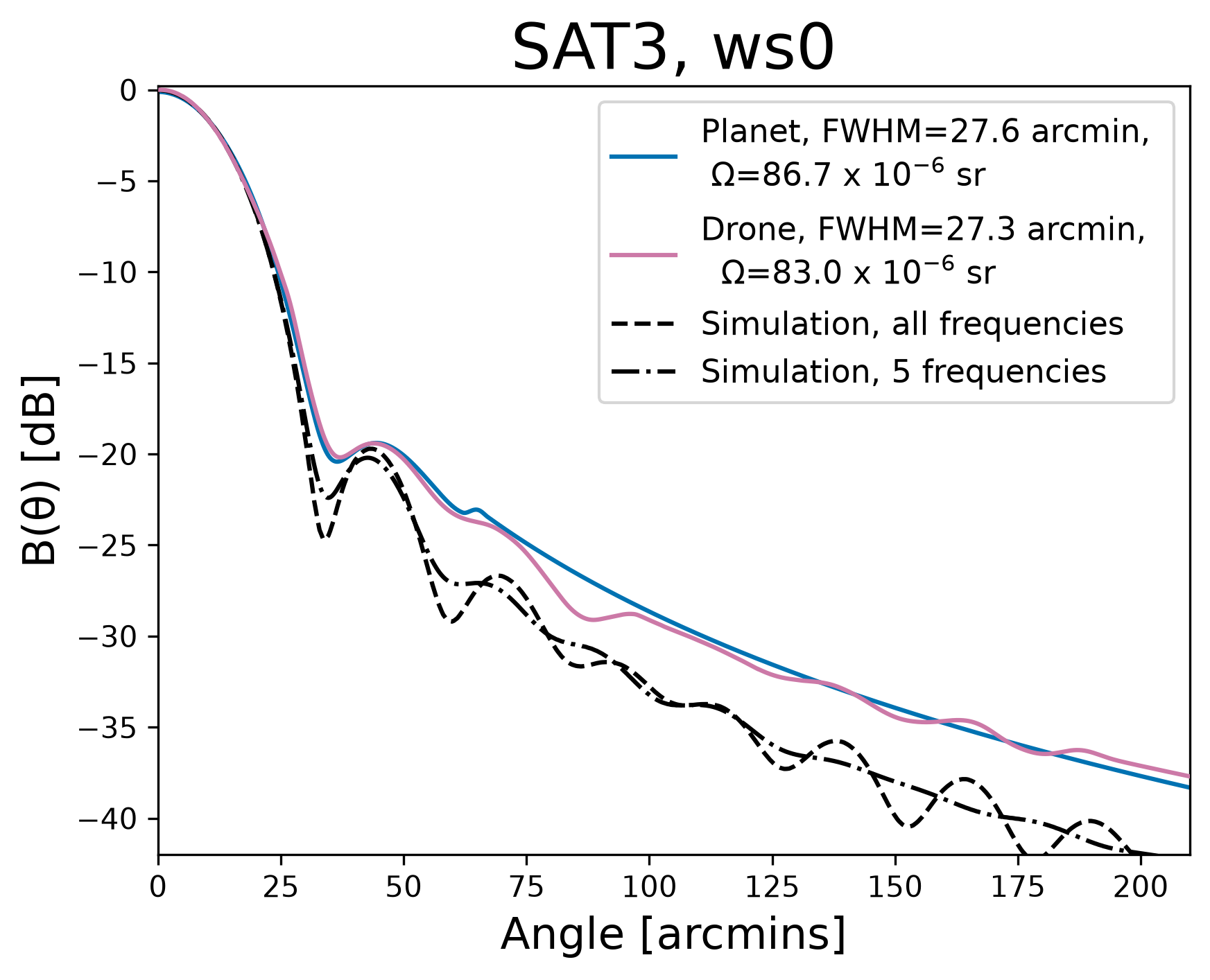}
    \caption{\label{fig:bandav_comparison_SAT1_and_SAT3}Top: A comparison of the band-averaged beam profiles for the planet (blue) and drone beams (green) for SAT1, ws0. The drone beam can preserve the data oscillatory structure at much larger angles compared to the planet beam. This is expected, since the drone beam is built from only a subset of frequencies, all with high transmission. Comparing the simulated profiles computed at the $\sim$\SI{1}{\giga\hertz} resolution of the bandpass (black dashed curves) with those computed from only the five frequencies measured with the drone (black dash-dotted curves) illustrates this effect. The amplitudes of the drone and Jupiter profiles agree for most of the studied angular range, yet both deviate from the simulated profiles after \SI{-3}{\decibel}. Bottom: The same comparison for SAT3, ws0. The drone measurements (red) deviate mostly from the planet measurement over the range most affected by SAT3's frequency-dependent feature. The drone profile again preserves the lobe structure until \SI{3}{\degree}, and is close in amplitude with the planet profile. Both profiles again show excess power compared to simulations.}

\end{figure}

The normalized, band-averaged radial profiles from planet, drone and simulations are shown in Figure \ref{fig:bandav_comparison_SAT1_and_SAT3} for SAT1, ws0 (top) and SAT3, ws0 (bottom), respectively. The simulation assumes a pixel at a \SI{3}{\centi\meter} radius from the boresight, corresponding to the average pixel radius of the ws0 wafers. Since the drone beams are measured one frequency at a time and each is normalized before averaging, the effective source weighting of the synthesized band-averaged beam is set purely by the Rayleigh–Jeans (RJ) bandpass; no color correction is therefore needed before comparing with Jupiter, which is itself very close to an RJ source \citep{planck_planetflux_2017}. Both SAT1 and SAT3  center wafer measurements show good consistency between the planet and drone beams main lobes; then the pronounced oscillatory structure of the frequency-resolved measurement takes over. The beam FWHM is slightly smaller for the drone, while the solid angles obtained by integrating the profiles up to \SI{3.5}{\degree} agree to within $6\%$ and $4\%$ for SAT1 and SAT3, respectively. This difference in FWHM is likely caused by the use of only a few selected frequencies together with the increasing slope of the preliminary bandpass, which acts as a weighting function in the construction of the band-averaged beam. As a result, the higher frequencies, which correspond to narrower beams, receive greater weight than the lower frequencies. However, the difference in amplitude, especially in the nearest lobes of SAT1, appears slightly larger than expected only from the difference in the measurement methods. This is mainly due to the nearest lobes at \SI{102}{\giga\hertz} (yellow dashed curve of Figure \ref{fig:radial_profiles_SAT1_and_SAT3}), which deviate from the amplitude scaling followed by the other frequencies.  We chose to retain this frequency in the average beam, since removing it would shift the effective frequency \SI{3}{\giga\hertz} lower, which has a larger impact on the small-angle part of the beam. For SAT3, the largest amplitude deviation shows over the angular range of the frequency-dependent feature. This also is well justified by having a subset of frequency measurements versus the full band measurement. 

\begin{figure}[H]
    \centering
    \includegraphics[width=\linewidth]{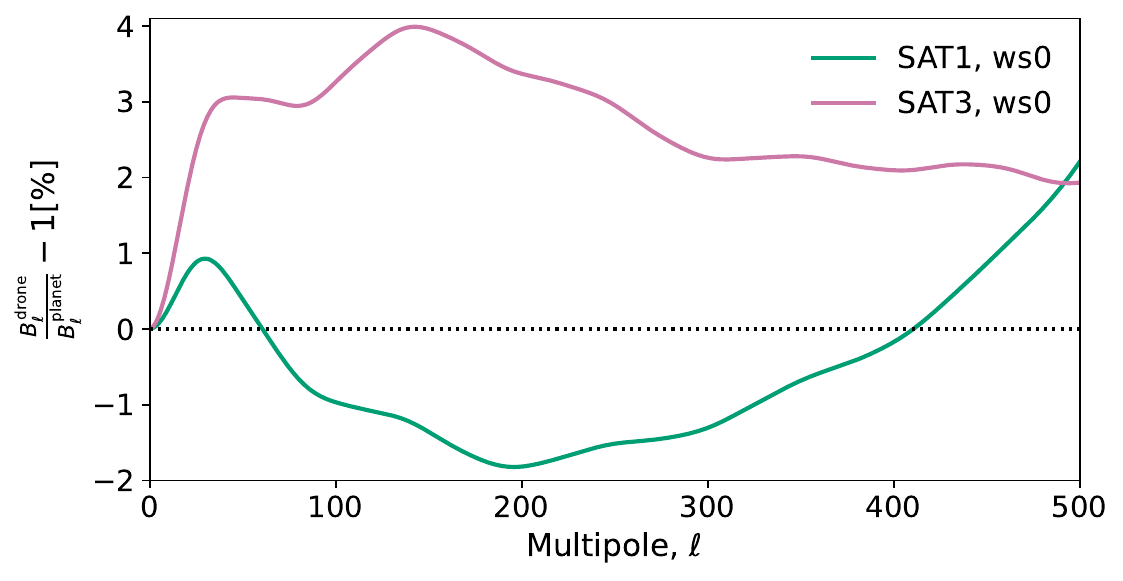}

    \caption{\label{fig:bandav_comparison_SATP1_and_SATP3_bell} The percent difference between the harmonic transfer functions of the drone and planet beams for SAT1, ws0 (green) and SAT3, ws0 (red). The error stays under $\sim 4\%$ for multipoles $\ell<500$.}
   
\end{figure} 

\noindent Any remaining amplitude mismatch in the main beam of both SAT1 and
SAT3 is expected to be mitigated after fitting and subtracting leakage templates built from
source-off measurements, which represents an ongoing effort. Preliminary results suggest that any leakage from the instrument and analysis sides is mostly consistent with a monopole whose amplitude is at the $1\%$ level of the source amplitude.

The drone beams, having significantly higher SNR than the Jupiter beams, allow for reconstructing the beam up to much larger radii before transitioning to the wing fit. For reference, the planet beam wing starts around $\sim$ \SI{60}{\arcminute} at all cases and around \SI{180}{\arcminute} for the drone. However, despite the more aggressive filtering required for Jupiter measurements, they better track the small-angle part of the beams. While the main beam is best characterized with planets and the far sidelobes can be measured from the Moon, the drone's advantage lies at the intermediate angular scales that bridge the two. More specifically, for both telescopes the drone profiles track the lobe centers over most of the angular range considered, except for the second lobe of SAT1, whose center is shifted to slightly larger angles than in the Jupiter profile again due to the deviating behavior of the \SI{102}{\giga\hertz} beam. Additionally, the drone measurements still show a prominent lobe structure at larger angles, qualitatively consistent with the simulated beams constructed from only the five measured frequencies (black dash-dotted curves). The simulated beams produced from all frequencies, in turn, motivate the model's transition to the $\sim\!1/\theta^{3}$ fit near \SI{120}{\arcminute}.

The planet and drone beam differences are further quantified through the percent difference in the harmonic transforms of their profiles \footnote{The methodology to derive the planet beam transfer functions will be described in \cite{Day-Weiss:2026}.} in Figure \ref{fig:bandav_comparison_SATP1_and_SATP3_bell}. From the figure, we establish again good consistency for the center wafers of both telescopes with the drone and planet beams having a maximum $\sim 4\%$ difference for multipoles up to $\ell=500$. Given the very sparse frequency sampling, the limited number of observations, and the different observing detectors (a third of the wafer for the drone versus the full wafer for the planet beam), the frequency-resolved calibration scheme appears to perform remarkably well for the studied wafers. In future campaigns we will sample a significantly wider frequency range with many more detectors, which we expect to reduce the drone–planet beam discrepancy substantially.
\section{Conclusion}
\label{sec:conclusion}

In this paper, we present a frequency-resolved beam calibration scheme using a narrow-band source emitting in the \SI{90}{\giga\hertz} band that is mounted on a drone and flown in the far field of the SO SATs, located in the Atacama Desert in Chile. These flights were performed during two campaigns designed as part of the POLOCALC \citep{Nati_2017} and HoverCal \citep{Dunner_2021} projects that aim to calibrate the beam and polarization angle of CMB telescopes in the \SI{90}{} and \SI{150}{\giga\hertz} bands to an accuracy of \SI{0.1}{\degree} or better. The measurements take place when the drone is at a fixed \SI{500}{\meter} distance from the telescopes and target the center wafers of both SAT1 and SAT3. The source frequency is tuned to five frequencies spanning a $20\%$ relative bandwidth around the center frequency of the SO MF1 band (\SI{93}{\giga\hertz}) for which the source beam was first characterized in the lab. Once the data are preprocessed, they are demodulated after retrieving the precise chopping frequency from both the telescope and source data. The source beam maps are then binned into azimuthally-averaged radial profiles and fitted using a model comprised of a Gaussian function for the core, a damped oscillation pattern for the nearest lobes and a modified $1/\theta^{3}$ scaling for the wing ($\geq$ \SI{3}{\degree}) as described in Section \ref{sec:beam_fitting_pipeline}. The results are presented both in terms of the measured beam scaling across the band in Section \ref{sec:single_f_beams} and through the comparison of their band-averaged product with the reconstructed planet beams which is shown in Section \ref{sec:planet_comparison}.

The single-frequency beam maps are re-centered and fitted with a 2D Gaussian model to compare the FWHM scaling against the frequency scaling expected from PO simulations. We find good overall consistency between the two, though both SAT1 and SAT3 exhibit a slightly shallower frequency dependence than the PO prediction, which serves as a useful reference rather than a definitive ground truth for the instrument. The fitted profiles of the per-frequency averaged maps follow the diffraction-limited optics expectation
for both telescopes, apart from a small deviation in the first two lobes of the SAT3 beams that is more prominent at higher frequencies (see Fig.~\ref{fig:radial_profiles_SAT1_and_SAT3}). Over this angular range we correct the parametric fit with an interpolating spline, which we allow to pass
through every point of the profile: the very high SNR near the beam center makes the map-to-map variance at a given frequency negligible. Since this distinct
SAT3 feature is well captured, we expect it to have no impact on the science if properly corrected for in the analysis.  Figure~\ref{fig:bells_continuous_SAT1_and_SAT3} shows the beam transforms of the fitted profiles, interpolated with splines at the SAT frequency
resolution between the lowest (\SI{84}{\giga\hertz}) and highest (\SI{102}{\giga\hertz}) measurements. For SAT1 we recover the expected FWHM scaling, captured by an $\alpha = 1.85 \pm 0.09$ correction in multipole space (see Eq.~\ref{eq:FWHM_scaling}). Truncating to $\ell \gtrsim 150$ yields $\alpha = 1.35 \pm 0.1$ for the SAT3 beam. 

The frequency choice and the validity of the frequency-resolved calibration method are assessed by comparison with the band-averaged beams reconstructed from Jupiter observations. Using preliminary estimates of the SAT bandpass, we synthesize the band-averaged beam from the drone observations. For this, we normalize all beams to unity at $\theta=0$, effectively dividing out both the source power and the detector gain. In future work we will subtract only the source power, so as to also build a continuous gain model across the band; this requires an end-to-end efficiency scheme that predicts the exact power expected on the detector for a given source power at each time, which remains an ongoing effort. Note that differences between the true bandpasses and the preliminary measurements used here could affect the consistency between the planet and drone beams; we plan to quantify this effect in future work. Figure~\ref{fig:bandav_comparison_SAT1_and_SAT3} shows the agreement between the drone and planet beams, which is strongest in the main beam for both SAT1 and SAT3; the remaining profile differences arise mainly because the drone beam is synthesized from only a subset of frequencies. The largest amplitude difference appears in the second lobe of the SAT1 beam, which is also slightly shifted outward. Both the different amplitude and center of the lobe are driven by the \SI{102}{\giga\hertz} beam which deviates from the overall scaling. For SAT3, the drone and planet profiles differ most over the \SIrange{40}{100}{\arcminute} range, where the frequency-dependent feature is strongest. The small residual main-lobe difference is consistent with the $1\%$ leakage measured when the same pipeline is applied to source-free observations. We expect this leakage to be suppressed with better crosslinking in future observations. The higher SNR of the drone beams also lets us extend the wing transition to much larger angles — almost three times larger than for the planet beams. For $\theta > \SI{2}{\degree}$, the drone data still show a pronounced lobe
structure, which we attribute to the sparse frequency grid: with only five highly transmissive frequencies entering the average, such features survive rather than wash out. This is reproduced when the average beam is reconstructed from only the five simulated beams at the measured frequencies (black dash-dotted curves in the same figure). Finally,
Figure~\ref{fig:bandav_comparison_SATP1_and_SATP3_bell} quantifies the drone–planet beam difference, showing a maximum error of only $\sim 4\%$ for
multipoles $\ell < 500$. Although the frequency coverage of these limited observations is sparse, and the detector set used for the drone covers only one third of the wafer compared to the full wafer used for the planet beam, the frequency-resolved calibration
scheme still performs remarkably well for the cases studied here. 

Having established the feasibility of frequency-resolved beam calibration for the SATs, we plan to extend the next campaigns both in the number
of measured frequencies and by scanning detectors at larger angles from boresight, aiming to characterize the beams down to \SI{-35}{\decibel} — better than is currently 
achievable with broadband sources \footnote{Figure~\ref{fig:bandav_comparison_SAT1_and_SAT3} shows that the planet beam becomes noise-dominated, forcing the switch to the $1/\theta^3$ fit, about 15\,dB higher than the drone beam.}. We are particularly interested in whether some frequencies carry more weight than others (frequencies at the turning points of the bandpass, for instance, may enable a continuous beam model in real space) and in whether the beam scaling varies across the focal plane. The analysis will be repeated for the \SI{150}{}, \SI{220}{} and \SI{280}{\giga\hertz} bands, using both POLOCALC and HoverCal data. The polarized maps will also be presented as part of future work, where we will set limits on instrument cross-polarization and explore any other non-azimuthally-symmetric features of the beam. Finally, we intend to apply the frequency-resolved scheme to polarization-angle studies as well, which will allow us to map the frequency-dependent behavior of the SATs HWPs
in the field. 

\section*{Acknowledgments}
This work was funded by grants from the Simons Foundation [MPS-Observatory-00457687, B.K.] and Simons Foundation International [SFI-MPS-Observatory-00007127, B.K.]. SO operates in the Parque Astronómico Atacama in northern Chile under the auspices of the Agencia Nacional de Investigación y Desarrollo (ANID). We thank the Republic of Chile for hosting SO in the northern Atacama, and the local indigenous Licanantay communities, whom we follow in observing and learning from the night sky.  This work would not be possible without the Simons Observatory site team based in Chile. 

We acknowledge the support of REUNA, AmLight, and Parque Astron\'omico in enabling the real-time transfer of data via a fiber-optic connection.  This research used resources of the National Energy Research Scientific Computing Center (NERSC), a Department of Energy User Facility (HEP project mp107 2023-2026).  The work presented in this article was performed on computational resources managed and supported by Princeton Research Computing, a consortium of groups including the Princeton Institute for Computational Science and Engineering (PICSciE) and Research Computing at Princeton University. Some of the software developed to produce this analysis can be found publicly available on the Simons Observatory GitHub organization (https://github.com/simonsobs).

ND, LB, FN, and AN acknowledge funding from the European Union (ERC, POLOCALC, 101096035). GC acknowledges funding from Italy Ministry of Research \\   (MSCA2024\_0000016). JCG acknowledges support from the Office of Science, Office of High Energy Physics, of the U.S. Department of Energy, under contract no. DE-AC02-05CH11231. CHC and RDP acknowledge ANID through BASAL CATA  FB210003. Work at Argonne National Laboratory was supported by the U.S. Department of Energy, Office of High Energy Physics, under contract DE-AC02-06CH11357. LP acknowledges support from the Wilkinson, Misrahi, and UC Louvain funds. YS acknowledges support from JSPS KAKENHI Grant Number 26KJ0059. 

This document was prepared by Simons Observatory
using the resources of the Fermi National Accelerator Laboratory (Fermilab), a U.S. Department of Energy, Office of Science, Office of High Energy Physics
HEP User Facility. Fermilab is managed by Fermi Forward Discovery Group, LLC, acting under Contract No.
89243024CSC000002.

\texttt{Claude} was used to refine selected parts of the analysis code, improve the formatting of a figure, and assist with language editing in certain parts of the manuscript. 

\bibliographystyle{mnras}
\bibliography{references.bib}{}

\appendix

\section{Flight statistics}
\label{flight_statistics}

The flight statistics for both campaigns are summarized in Table \ref{tab:flight_stats}, including those not used in the present analysis and for which results will be reported in future publications. We distinguish between the total number of accumulated flights and the number of usable flights for each combination of telescope, wafer, and source frequency. Note this table only refers to flights dedicated to calibration and does not include the ones performed for testing the stability of the gimbal. A flight is declared `non-usable' if a large group of detectors ($>$ $100$) are saturated. As the system power is stable within $\sim$ $1-2$ \SI{}{\dBm}, the saturation is the result of the atmospheric loading that changes non-negligibly during the day. If a flight needed to be aborted early, it is not included at all in the table. Note that the number of usable flights for the \SI{93}{\giga\hertz} measurements is different than the one quoted in Figure \ref{fig:93_ghz_maps}
and the corresponding panels of Figure \ref{fig:monochromatic_avg_maps} since the figures only include the flights performed for the middle wafer part for consistency with the other frequencies.

\begin{table}[H]
    \centering
    
    \begin{tabular}{|c|c|c|c|}
    \hline
       Telescope & Wafer & Frequency (\SI{}{\giga\hertz}) & Usable / Total flights \\

       \hline
       SAT1 & ws0 & 84 & 7/8 \\

       \hline
       SAT1 & ws0 & 87 & 2/2 \\

       \hline
       SAT1 & ws0 & 93 & 18/24  \\

       \hline
       SAT1 & ws0 & 99 & 2/2 \\

       \hline
       SAT1 & ws0 & 102 & 7/8 \\

       \hline
       SAT1 & ws2 & 93 & 4/7 \\

       \hline
       SAT3  & ws0 & 84 & 8/9 \\

       \hline
       SAT3  & ws0 & 87 & 3/3 \\

       \hline
       SAT3  & ws0 & 93 & 24/29 \\

       \hline
       SAT3  & ws0 & 99 & 3/3 \\

       \hline
       SAT3  & ws0 & 102 & 8/9 \\

       \hline
       SAT3  & ws6 & 93 & 12/14 \\

       \hline
       SAT3  & ws3 & 84 & 3/3 \\

       \hline
       SAT3  & ws3 & 87 & 3/3 \\

       \hline
       SAT3  & ws3 & 93 & 9/11 \\

       \hline
       SAT3  & ws3 & 99 & 3/3 \\

       \hline
       SAT3  & ws3 & 102 & 2/3 \\

       \hline
         
    \end{tabular}
    \caption{\label{tab:flight_stats}Total count of useful flights out of the total calibration flights accumulated for each telescope, wafer and source frequency for both April and December 2025 campaigns.}
\end{table}

\section{Single-frequency beam maps}
\label{single_f_beam_maps}
\begin{figure}
    \centering
    \includegraphics[width=0.49\linewidth]{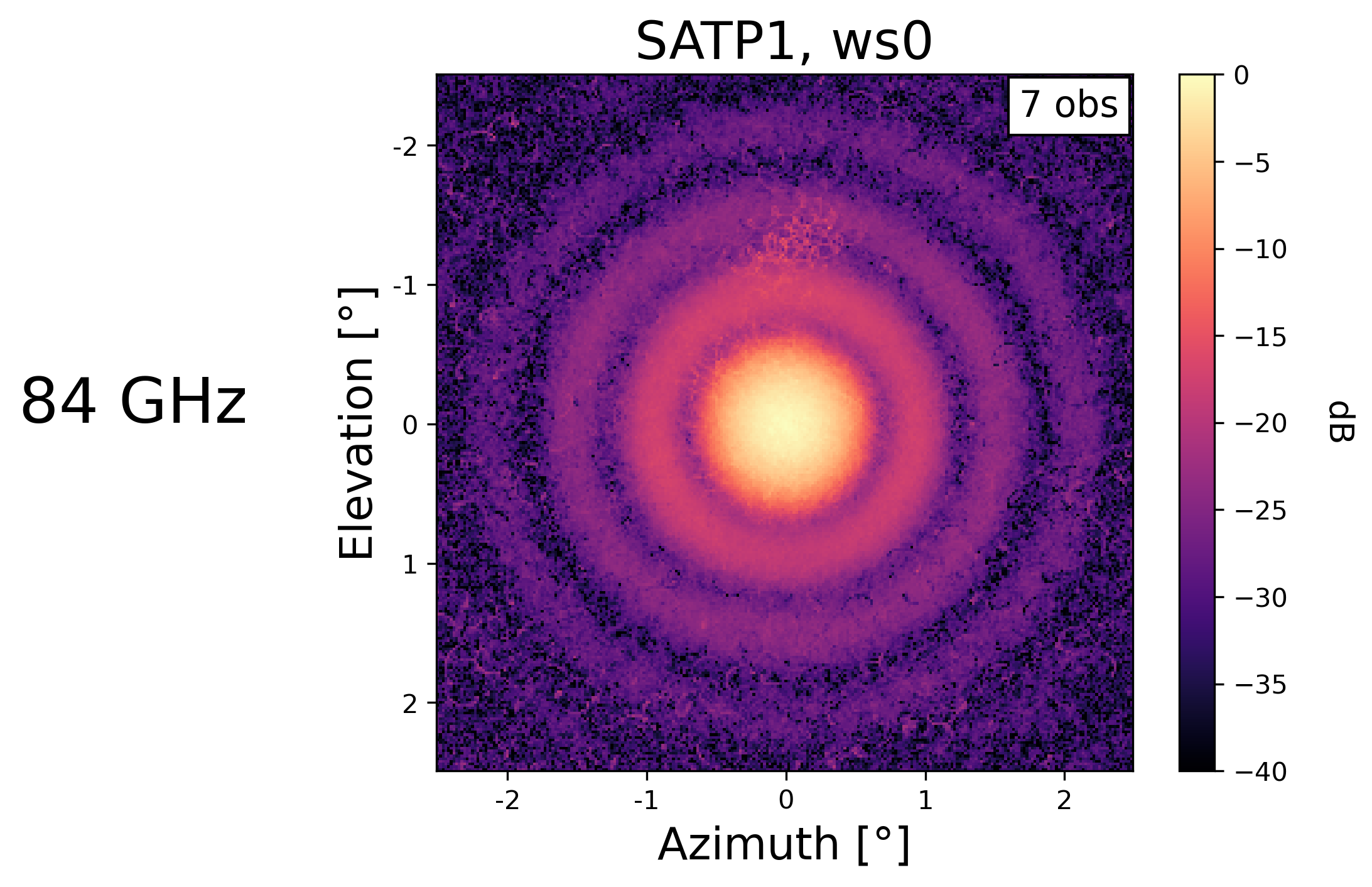}\includegraphics[width=0.38\linewidth]{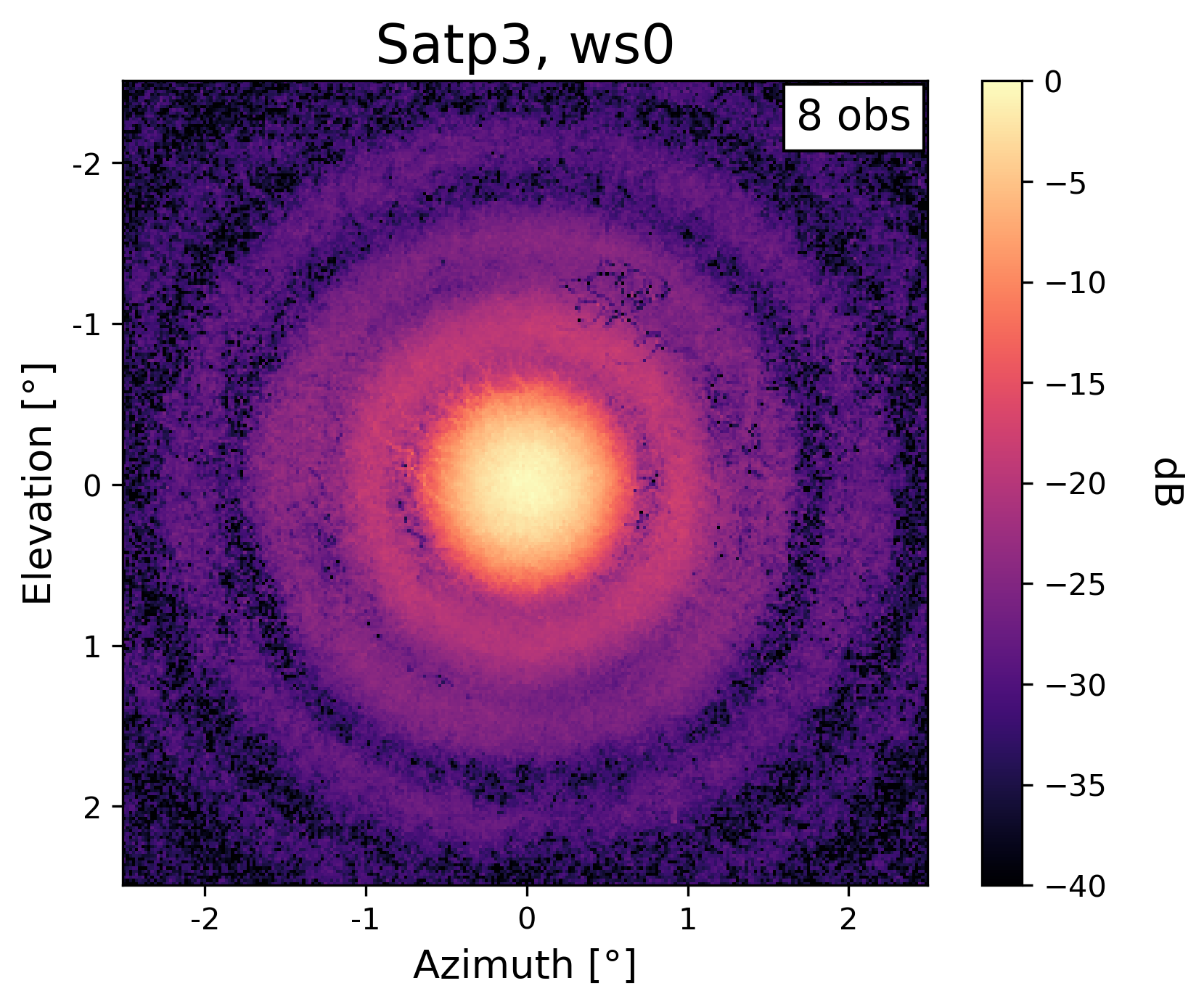}

    \includegraphics[width=0.49\linewidth]{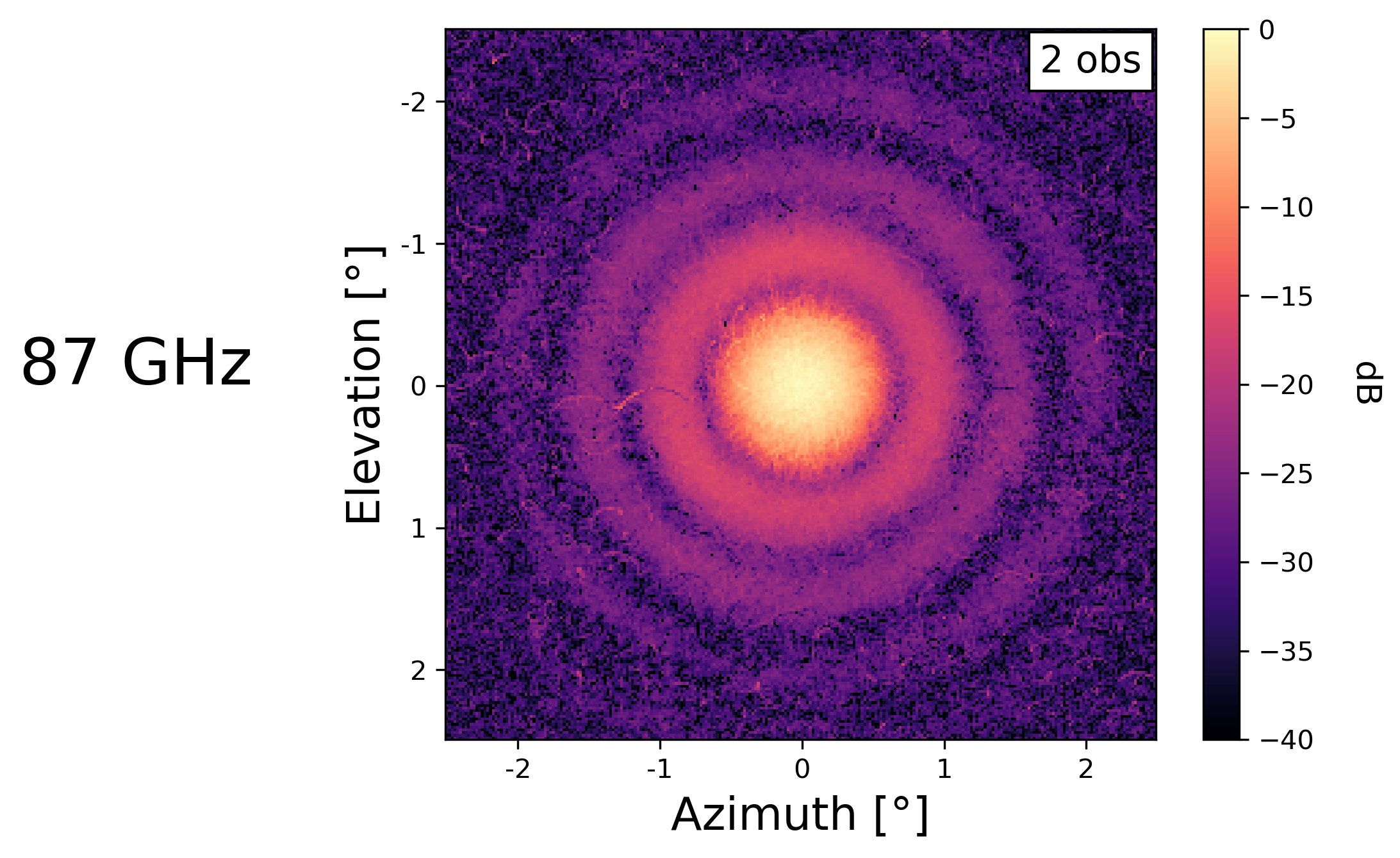}\includegraphics[width=0.38\linewidth]{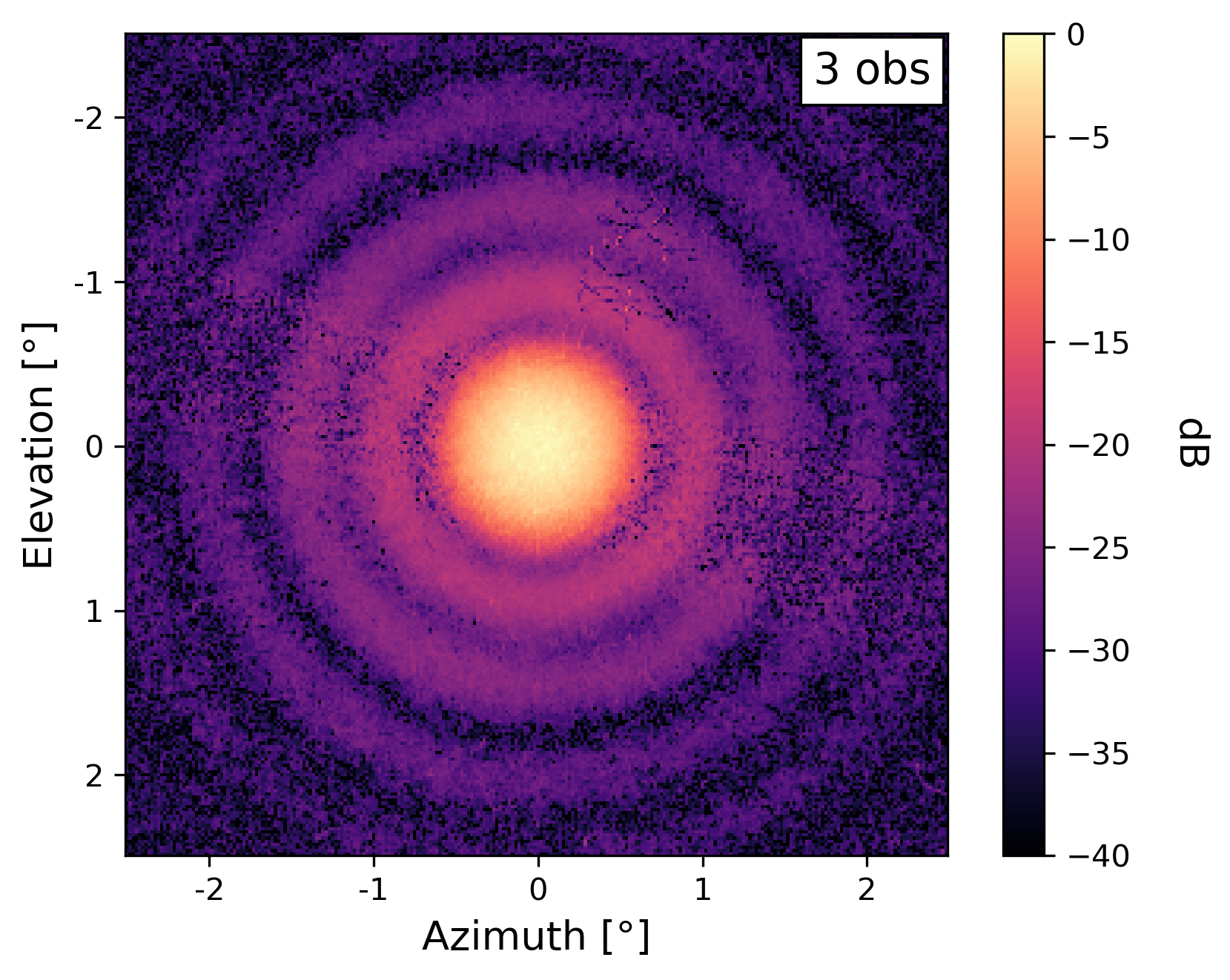}

    \includegraphics[width=0.49\linewidth]{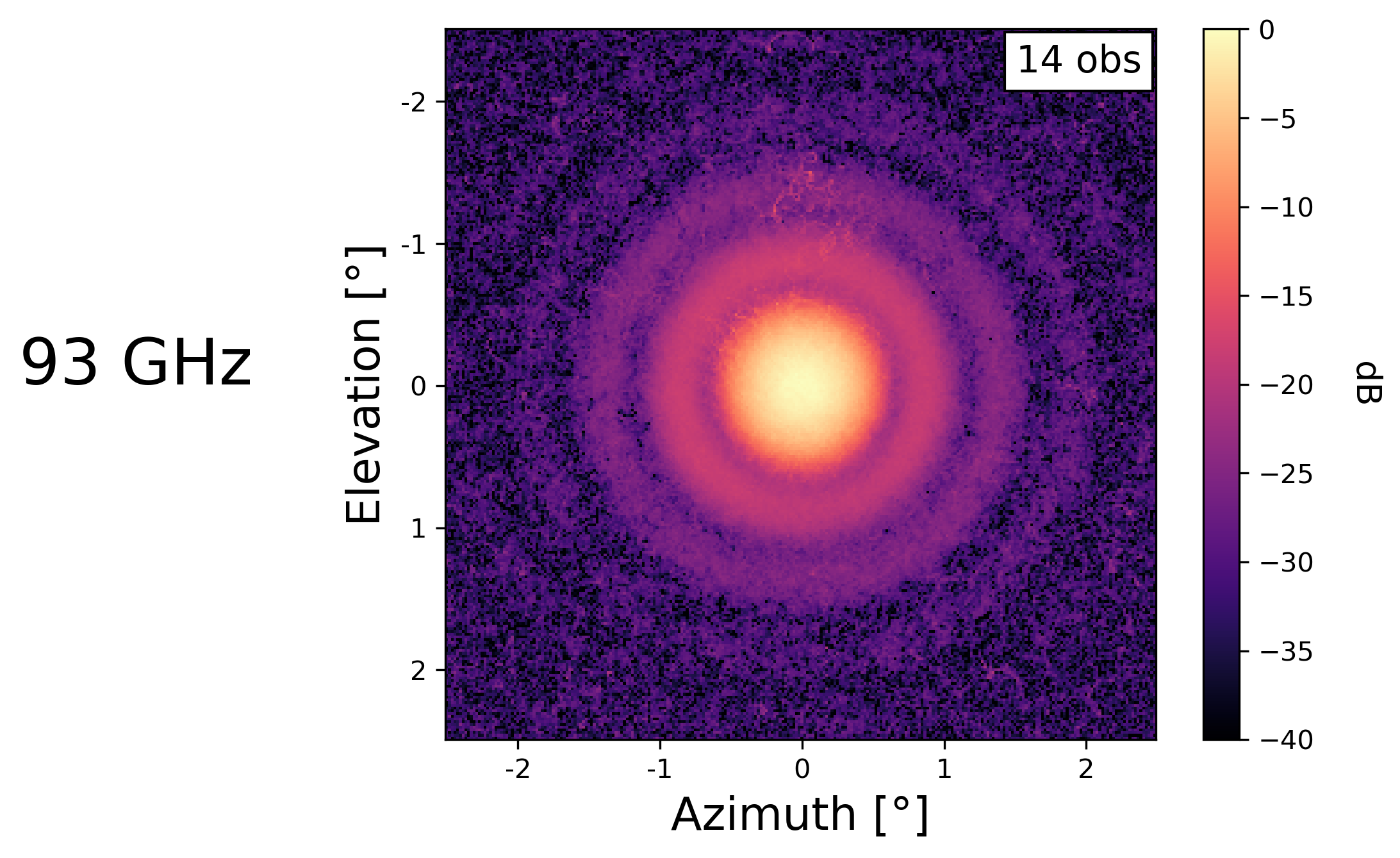}\includegraphics[width=0.38\linewidth]{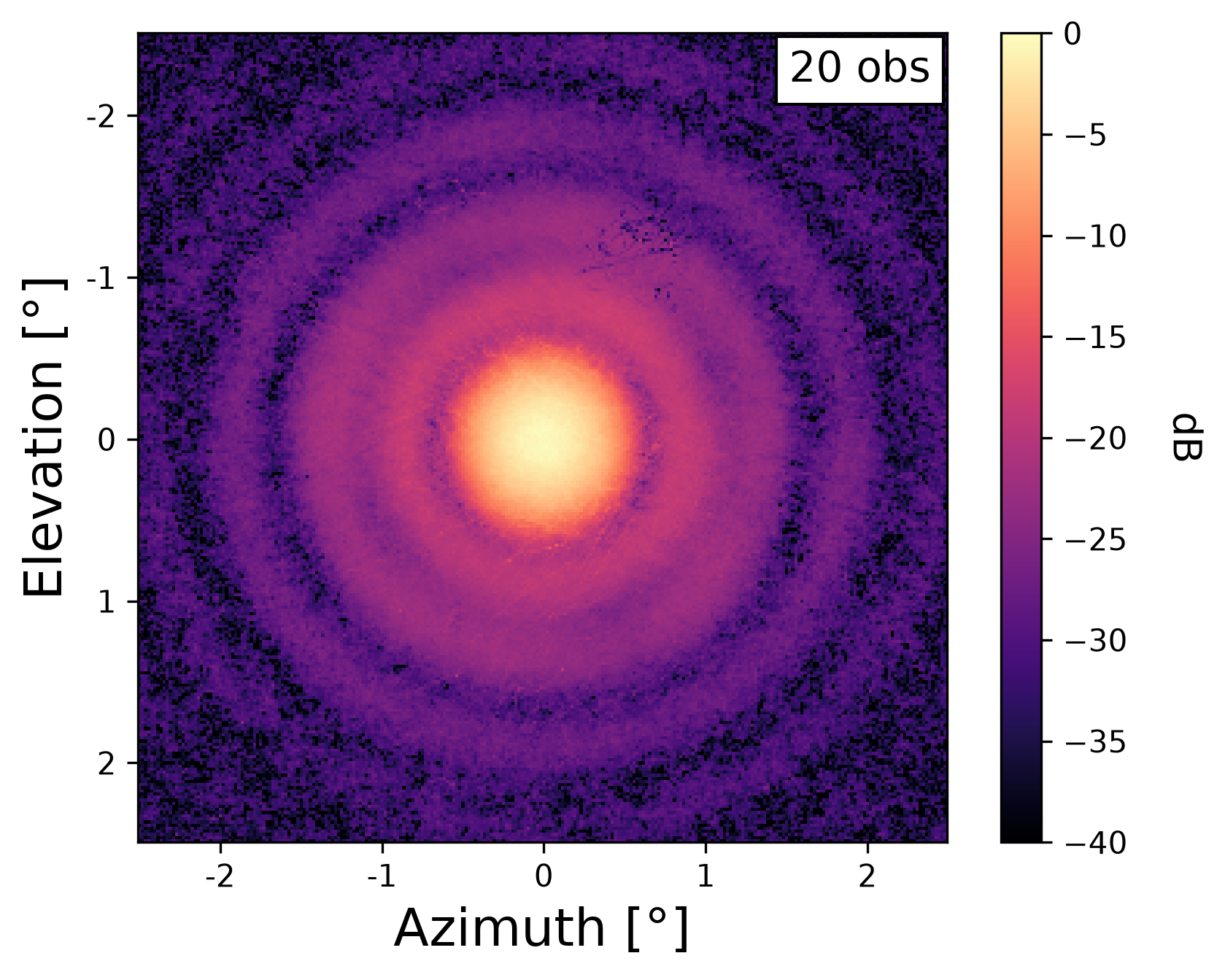}

    \includegraphics[width=0.49\linewidth]{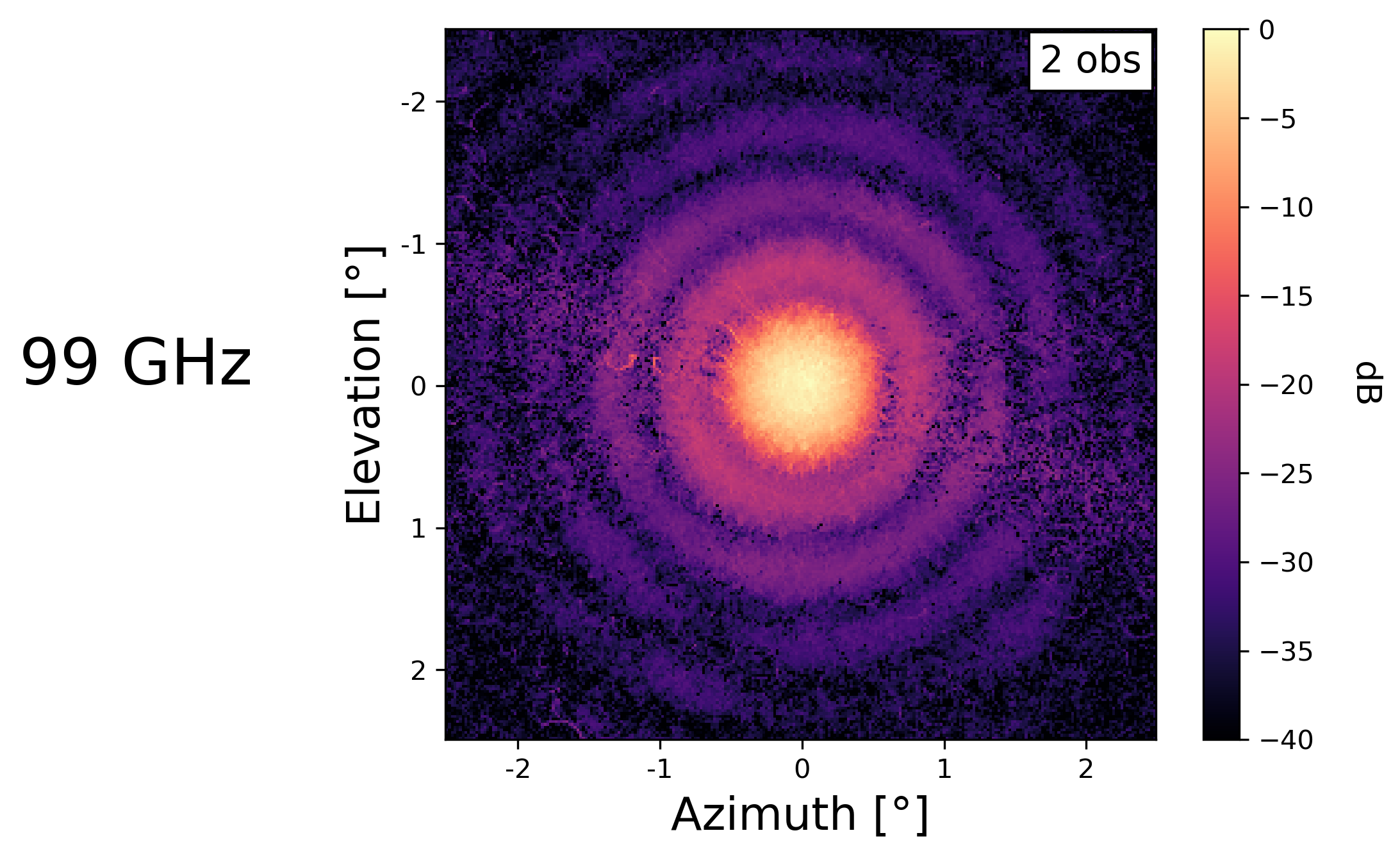}\includegraphics[width=0.38\linewidth]{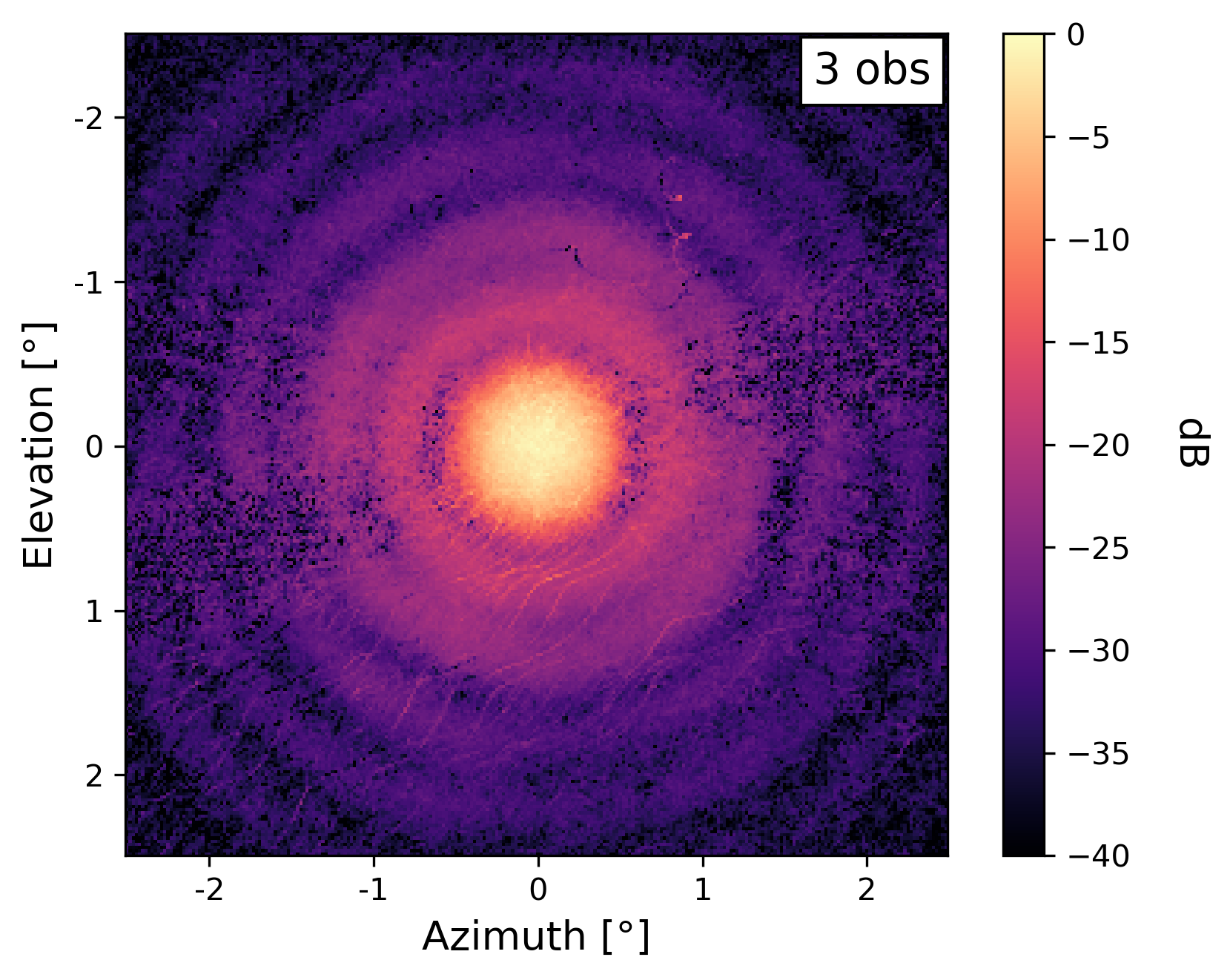}

    \includegraphics[width=0.49\linewidth]{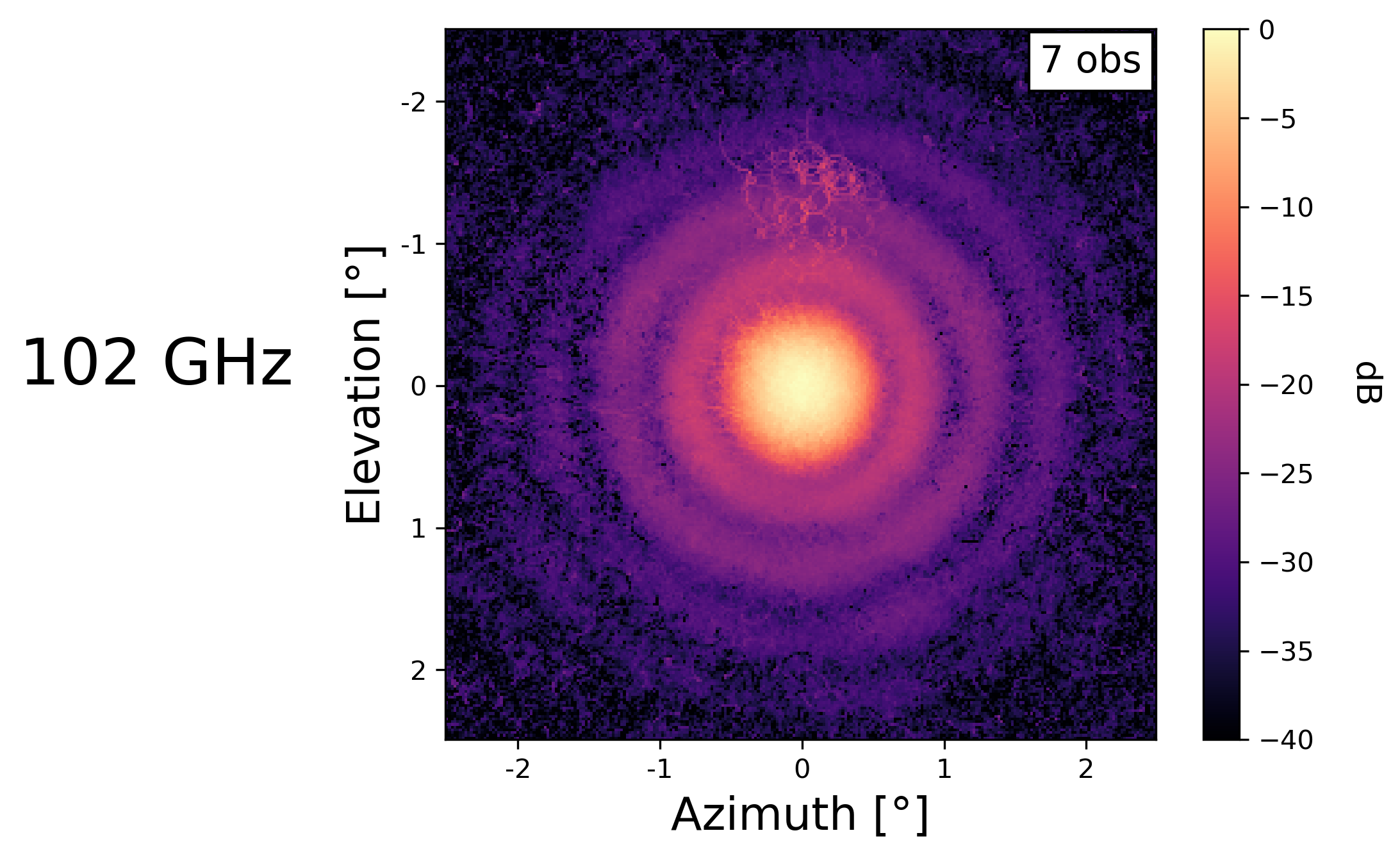}\includegraphics[width=0.38\linewidth]{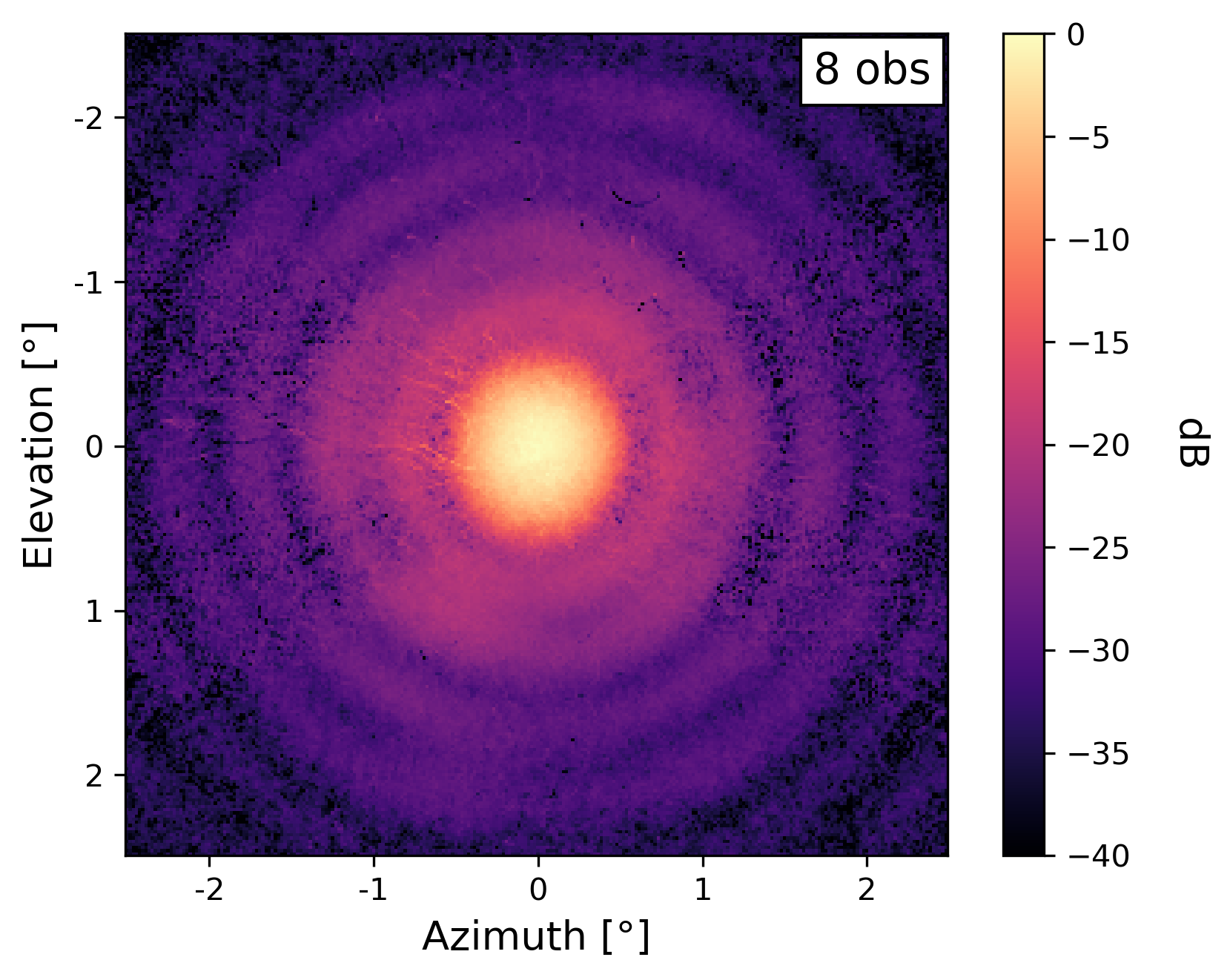}

        \caption{\label{fig:monochromatic_avg_maps} The inverse-variance weighted average maps for 84, 87, 93, 99 and \SI{102}{\giga\hertz} (rows 1-5) of SAT1, ws0 (left), and SAT3, ws0 (right). As frequency increases, the size of the beam consistently decreases across both cases.}
\end{figure}

This section presents the inverse-variance-weighted average maps for each frequency. The left and right columns of Figure \ref{fig:monochromatic_avg_maps} correspond to SAT1, ws0 and SAT3, ws0, respectively. Rows 1–5 show frequencies ranging from \SI{84}{} to \SI{102}{\giga\hertz}. A clear decrease in beam size is visible from the top to the bottom of the figure, consistent with the increasing frequency. In addition, the first and second sidelobe rings of the SAT3  beams become increasingly difficult to distinguish after $\nu\geq 93$ \SI{}{\giga\hertz}. Finally, a faint X-shaped pattern is visible, particularly in the higher-frequency maps. This feature is most likely caused by leakage from the thermal emission of the drone body, a fraction of which inevitably leaks to demodulated signal. The source is modulated at a $\sim$\,\SI{e-5}{\hertz} band around $f_{\mathrm{chop}} = \SI{37.00015}{\hertz}$, while the demodulated data are low-pass filtered at \SI{2}{\hertz}. Applying a substantially lower cutoff frequency could help suppress this contamination but would also introduce significant ringing into the data.

\section{SAT3  parametric fitting residuals}
\label{SAT3_sl_bump_vs_f}
\begin{figure}[h!]
    \centering
    \includegraphics[width=0.7\linewidth]{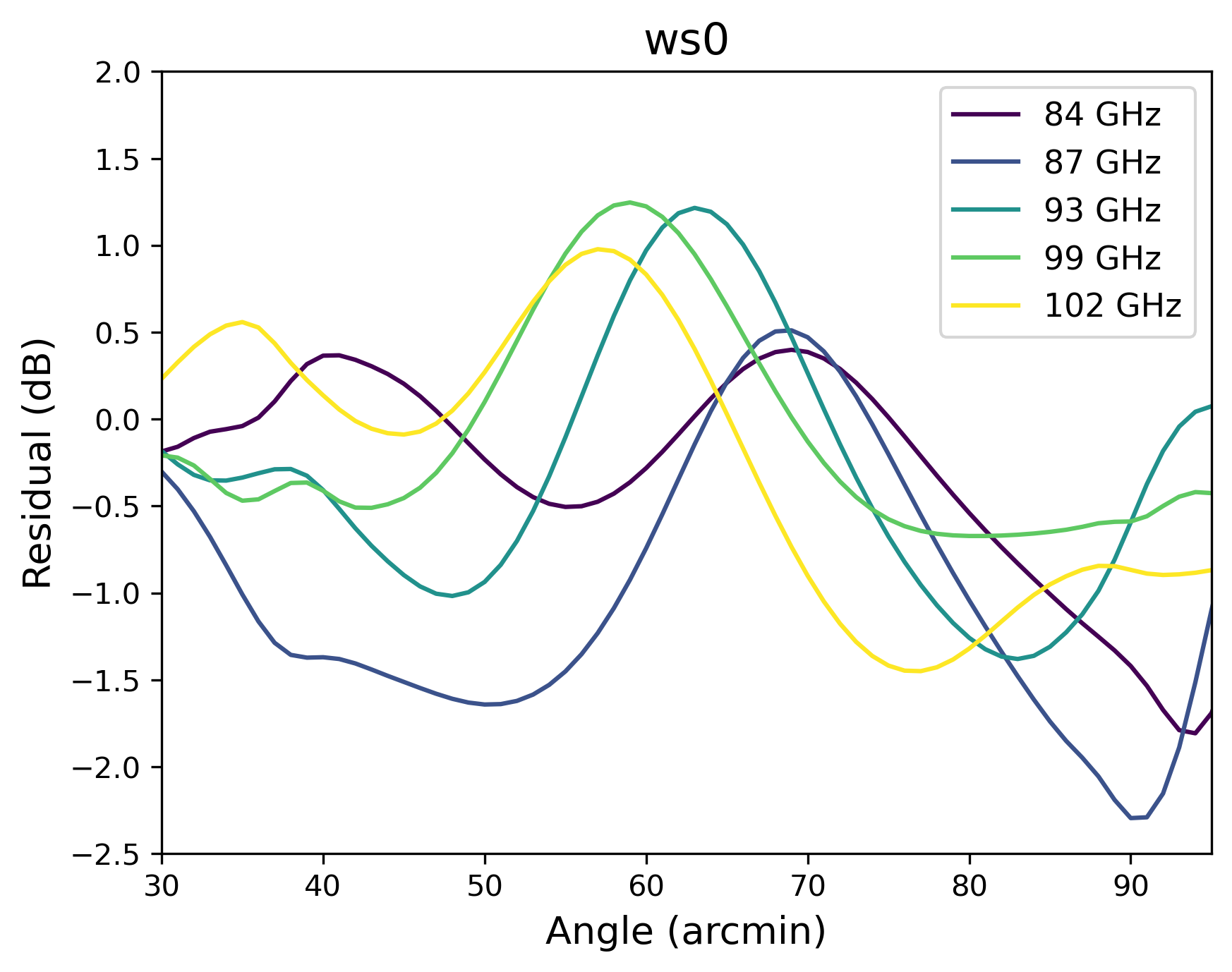}

    \caption{\label{fig:residual_param_bias_vs_f} Average residual bias between data and parametric model fits per frequency after smoothing with a Gaussian kernel for SAT3, ws0. The residual, Gaussian-looking structure moves to smaller angles as frequency increases while from \SI{84}{} to \SI{99}{\giga\hertz,} the amplitude also increases consistently with frequency.}

\end{figure}

The parametric model discussed in Section \ref{sec:beam_fitting_pipeline} describes adequately the SAT1 data but fails to adapt to the intricate structure of the first two lobes of the SAT3  beam profiles. The residual bias appears as a Gaussian-looking `bump' centered around 2.5 times the beam FWHM and increases in amplitude with increasing frequency. An exception to this statement is again the \SI{102}{\giga\hertz} case, as already expected from the results presented in Section \ref{sec:single_f_beams}. Figure \ref{fig:residual_param_bias_vs_f} shows the average residual between data and parametric model fits per frequency, around the angular range over which we applied the spline correction for SAT3, ws0. The residuals are smoothed with a Gaussian kernel with $\sigma$=\SI{1.5}{\arcminute} for visualization purposes. 

Because the near-lobe effect is more pronounced in the \SI{150}{\giga\hertz} band \citep{Day-Weiss:2026}, we expect to model it more accurately by repeating the chromaticity analysis at that frequency band. We plan to obtain the relevant measurements in the next campaign (currently planned for December 2026). If a more precise characterization of the frequency dependence still does not reveal the origin of the effect, and since this issue appears only in SAT3, we could consider different interventions. While we strongly believe the AR coatings to be the cause of this feature, one could consider alternative tests such as temporary blackening of the forebaffle or a brief forebaffle swap to assess the impact on the beams.

\end{document}